\documentclass[11pt]{article}
\usepackage{natbib}
\usepackage[hidelinks]{hyperref}
\usepackage{amsmath,amssymb,amsfonts,amsthm,mathtools,latexsym}
\usepackage[all]{xy}
\usepackage{float}
\usepackage{graphicx,epstopdf}
\usepackage{url}
\usepackage{appendix}
\usepackage{enumerate}
\usepackage{multirow}
\usepackage{xcolor}
\usepackage[left=2cm,top=2cm,right=2cm,bottom=2cm]{geometry}
\usepackage{booktabs}
\usepackage{hyperref}
\usepackage{caption}
\numberwithin{equation}{section}

\title{A stochastic dose-response framework\\
for environmentally persistent pathogens}
\author{
Mahmudul Bari Hridoy$^{1,2}$, 
Arik Hartmann$^3$, 
Kate E. Langwig$^3$, \\ 
Joseph R. Hoyt$^3$,  
Lauren M. Childs$^{1,2}$\\ 
~\\$^1$Department of Mathematics, Virginia Tech\\
$^2$Center for the Mathematics of Biosystems, Virginia Tech\\
$^3$Department of Biological Sciences, Virginia Tech\\}
\date{}

\begin{document}

\maketitle

\vspace{0.25cm}

\begin{abstract}

Infectious diseases caused by environmentally persistent pathogens can strongly affect host populations because transmission occurs not only through direct host-host contact but also via indirect exposure to contaminated environments. While in some systems environmental reservoirs help sustain exposure even when infected host numbers are low,  reliance on an environmental transmission pathway may also increase pathogen extinction risk. Infection risk may depend on both pathogen dose and the timing of exposure. Thus, understanding how dose--response, stochasticity, and seasonal changes in host susceptibility or contact rates shape pathogen invasion and persistence remains an important challenge for environmentally transmitted disease systems. To address this, we develop a stochastic dose--response framework that integrates host infection dynamics with an explicit environmental pathogen reservoir. Transmission occurs through both direct contact with infectious hosts and indirect environmental exposure, with infection probability governed by dose-response functions. We focus on a stochastic continuous-time Markov chain formulation and use branching process approximations to estimate disease extinction probabilities when infected hosts or environmental pathogen loads are low. We then extend the framework to include seasonality in host susceptibility, environmental contact, and host--host contact.  As a case study, we apply the model to snake fungal disease. Numerical simulations show that dose--response influences epidemic takeoff and infection levels, while seasonality creates windows of high and low extinction risk. These extinction risks depend strongly on the route and timing of introduction within the seasonal cycle. These results, coupled with global sensitivity analysis, illustrate how stochasticity, nonlinear dose--response, and seasonal timing shape outbreak dynamics for environmentally persistent pathogens.

\end{abstract}




\section{Introduction}

Infectious diseases can severely affect host populations, particularly when pathogens persist in the environment and alter infection risk through indirect environmental exposure, even when infected host numbers are low \cite{hoyt2020environmental, lanzas2020modelling}. As a result, pathogen transmission depends not only on direct host-host contact but also on indirect exposure through contaminated environments, generating substantial heterogeneity in the timing and magnitude of pathogen exposure that can sustain transmission and complicate outbreak control \cite{de2005mechanisms, rohani2009environmental}. Whether an exposure event leads to successful transmission depends on several key determinants, including the number and timing of contacts across transmission routes and the dose of pathogen encountered. Combining  environmental pathogen dynamics with host dynamics links empirical studies of pathogen exposure to mechanistic theory of disease spread, allowing us to evaluate the relative contributions of direct and indirect transmission.

In many disease systems involving environmentally persistent pathogens, environmental pathogen survival and variation in host–environment contact can complicate understanding of transmission dynamics \cite{rohani2009environmental,breban2013role}. For pathogens capable of surviving for long periods outside the host, such as \textit{Vibrio cholerae} in water systems, \textit{Bacillus anthracis} spores in soil, or \textit{Clostridioides difficile} on hospital surfaces, the environment functions as an active reservoir that can sustain transmission over time \cite{reidl2002vibrio, carlson2018spores, borji2022prevalence}. After shedding, a process where pathogens are released from infected hosts into the environment, pathogens may persist outside the host for extended periods, with survival and transmission efficiency shaped by environmental conditions such as temperature, moisture, and pH. This generates substantial variation in exposure risk \cite{breban2013role}. Importantly, environmental persistence allows transmission to continue even when host infection prevalence is low. As a result, explicit modeling of pathogen reservoirs and host exposure processes is essential for understanding transmission dynamics in environmentally mediated disease systems.

Classical mathematical models of infectious disease transmission primarily focused on deterministic models with direct host–to-host contact, with environmental effects typically incorporated implicitly rather than through explicit pathogen reservoirs as dynamical components \citep{anderson1979population}. Later extensions introduced environmental pathogen reservoirs as explicit state variables, motivated in part by environmentally persistent pathogens \citep{breban2013role,eisenberg1996quantifying,eisenberg2002disease,li2009dynamics}. Much of this early work was motivated by human pathogens, particularly waterborne diseases such as cholera, where pathogen persistence in the environment plays a central role in transmission \citep{codecco2001endemic,king2008inapparent,tien2010multiple}. Since the early work, deterministic models for environmentally transmitted infectious diseases (ETIDs) have been extended to a range of settings, including systems with multiple transmission pathways and environmental persistence (see, e.g., \cite{, tien2010multiple, bani2012effectiveness, breban2010general, gautam2011modeling}). Contact patterns, pathogen dose, and interactions with environmental conditions can differ substantially between host–host and environmental transmission, giving environmental reservoirs a unique capacity to shape transmission and persistence. Thus, even when environmental reservoirs are incorporated mathematically in a manner similar to an additional host, their effects on transmission dynamics are often distinct. 

When a pathogen persists in the environment, the initial trigger for an outbreak is often a rare, isolated event \cite{engering2013pathogen}. While deterministic approaches capture average patterns of transmission, transmission itself is inherently random. The stochasticity of transmission events is especially apparent during the early stages of invasion when the number of infectious hosts or level of pathogen in the environment is low. At this stage, deterministic thresholds based on average behavior may provide an incomplete description of invasion risk \cite{allen2012extinction,hridoy2025investigating}. In particular, they cannot quantify the probability that a pathogen introduced at low abundance will go extinct rather than persist, despite threshold quantities such as the basic reproduction number exceeding one \cite{allen2012extinction}. Thus, extending deterministic environmental transmission models to stochastic frameworks enables us to quantify extinction probabilities and capture early outbreak dynamics.

While stochastic epidemic models are used to study the transmission of ETIDs \cite{zhao2019analysis,wang2012simple, shakiba2021effects,lahodny2015estimating}, many existing stochastic ETID models either neglect indirect transmission through the environment or do not explicitly incorporate the replication of free-living pathogens.  A counter example to this is Lahodny et al. \cite{lahodny2015estimating}, who use a continuous-time Markov Chain (CTMC) to show the importance of particular parameters for invasion risk. However, they assumed a fixed functional relationship between environmental pathogen load and infection and did not consider periodically varying parameters.

Seasonal variation can strongly influence environmentally mediated pathogen transmission by altering both host susceptibility and pathogen viability in the environment  \cite{dowell2001seasonal, polgreen2018infectious}. Changes in temperature, humidity, rainfall, and sunlight affect how long pathogens remain viable in the environment, while also influencing host behavior and contact with contaminated environments \cite{walther2004pathogen}. Together, these processes shape transmission rates and the timing and intensity of outbreaks. As a result, disease dynamics in seasonally varying environments can differ substantially from those predicted by models with constant parameters, making it more difficult to predict disease emergence and epidemic trajectories. Explicitly accounting for seasonal variation in environmental persistence and host–environment contact is therefore important for understanding transmission in environmentally mediated disease systems.

In environmentally mediated systems, infection is not determined solely by contact rates but by how exposure translates into successful transmission. Therefore, the functional relationship between pathogen dose and infection probability  plays a central role in shaping transmission dynamics. Brouwer et al. \cite{brouwer2017dose} showed that the choice of dose–response function plays an important role and different functional forms that are consistent with experimental data can lead to qualitatively different transmission dynamics and outbreak predictions, particularly at low doses. Despite the importance of environmental transmission across many
infectious diseases, the impact of dose--response structure on epidemic dynamics, particularly in seasonal environments, remains largely unexplored in mathematical models. Consequently, examining how pathogen dose is represented and carefully considering its functional form in both deterministic and stochastic modeling frameworks is an important step toward understanding environmentally mediated transmission.

To address these challenges, we develop a stochastic modeling framework using a continuous-time Markov chain (CTMC) that integrates host infection dynamics with an explicit environmental pathogen reservoir. Transmission occurs through direct contact with infectious hosts and indirect exposure to pathogens in the environment, with infection probability governed by dose–response functions that link environmental pathogen load to infection risk and capture variability in pathogen exposure. This formulation captures transmission dynamics when host or pathogen densities are low and allows infection risk to depend on the prevailing environmental pathogen load as well as on direct contact events between hosts.
We formulate the CTMC as a stochastic extension of a corresponding nonlinear system of ordinary differential equations (ODE), with time non-homogeneous transition rates to define the infinitesimal transition probabilities. Using branching process approximations, we derive outbreak probabilities from the underlying CTMC. We further extend this framework to incorporate seasonality in host susceptibility and environmental exposure, reflecting empirically documented seasonal patterns in environmentally transmitted diseases. This allows us to examine how the timing and relative strength of direct and environmentally mediated transmission, including their synchrony or asynchrony, shape invasion, persistence, and extinction dynamics.

As a case study, we apply our framework to snake fungal disease (SFD), an emerging wildlife disease that has caused widespread and severe infections in wild snake populations across North America \cite{conley2025mortality}. As a result, concerns about population-level impacts demonstrate the need for a mechanistic understanding of its transmission dynamics \cite{lorch2016snake}. The environmentally persistent nature of \emph{Ophidiomyces ophiodiicola} \cite{campbell2021soil}, combined with seasonal variation in snake behavior and susceptibility \cite{lind2023ophidiomycosis}, makes snake fungal disease an ideal system for examining how dose-dependent environmental exposure and stochasticity jointly shape invasion and extinction dynamics.

We first detail the modeling framework in Section~\ref{sec:model_formulation} focusing on our dose-response and seasonally varying functions. In Section~\ref{sec:analytic_results}, we present the analytical results, including derivation of the branching process approximation and outcome measures. Next, in Section \ref{sec:num_results}, we use numerical simulations to examine how dose--response, seasonal timing, and route of introduction affect outbreak dynamics as well as assess the importance of parameters with global sensitivity analysis. We conclude with a discussion of the implications, limitations, and future directions of the framework in Section~\ref{sec:discussion}.

\section{Model formulation} \label{sec:model_formulation}

We consider a single host species with susceptible and infected classes, denoted by $S$ and $I$, respectively, and an environmental pathogen reservoir, $W$. 
Recruitment into the host population occurs at rate $\pi$ and loss occurs with natural mortality at rate $\mu$. 
Infected individuals recover at rate $\phi$, die from disease at rate $\delta$, and shed pathogen into the environment at rate $\alpha$. 
Environmental pathogen dynamics include removal at rate $\nu$. Transmission occurs through exposure to pathogen in the environment and through direct contact with infectious hosts, modeled via dose-response functions $f_W$ and $f_I$, respectively. Environmental contact occurs at rate $\kappa(t)$ with environmental exposure scaled by $p$ and direct host--host contact at rate $c(t)$ with host infectivity scaled by $b$. Throughout, we assume a linear dose-response function for direct transmission ($f_I$), so that direct transmission is proportional to the current host prevalence
$I(t)$. Environmental transmission is mediated by the pathogen load $W$ and
follows one of three potential dose-response functions (Eqs.~\eqref{eq:linear}--\eqref{eq:Beta_pois_approx}), described in detail below. Variation in host or seasonal factors that alter transmission is captured by a time--dependent relative susceptibility factor $\gamma(t)$.  The resulting deterministic model is

\begin{align}\label{eq:SIW}
\dot S &= \pi (S+I)
- \gamma(t)\Big(\kappa(t)\, f_W(p W) + c(t)\, f_I(b I)\Big) S
+ \phi I - \mu S, \nonumber \\[4pt]
\dot I &= \gamma(t)\Big(\kappa(t)\, f_W(p W) + c(t)\, f_I(b  I)\Big) S
- (\phi + \mu + \delta)\, I, \nonumber\\[4pt]
\dot W &= \alpha I - \nu  W .
\end{align}

\begin{figure}[H]
\centering
\includegraphics[width=.6\textwidth]{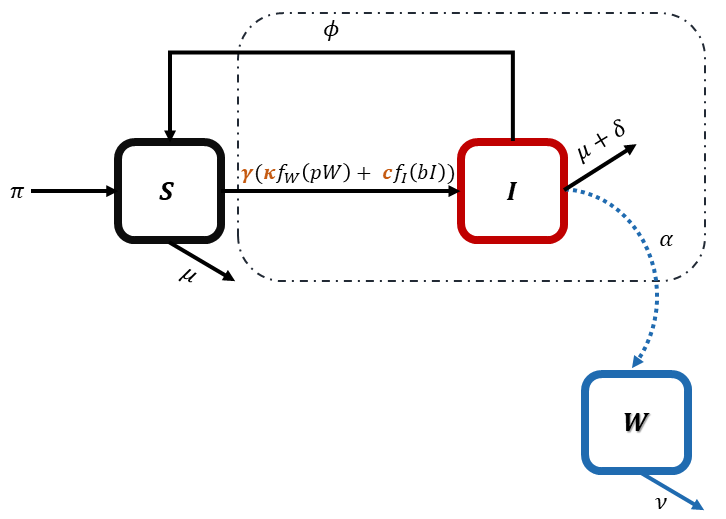} 
\caption{Compartmental diagram of the model. States include: $S$, susceptible hosts; $I$, infectious hosts; and $W$, pathogen in the environment. Solid black arrows indicate transition of hosts between states. Solid blue arrows indicate pathogen entrance and removal from the environment. Parameter descriptions are found in Table \ref{ParamTable}. Parameters shown in orange denote time–periodic rates. The dashed–dotted curve encircles stage $I$ and the rates associated with the branching process approximation. }\label{fig:Single_SFD_comp}
\end{figure}

The system is shown in a compartmental diagram in Fig.~\ref{fig:Single_SFD_comp}, with model parameters and descriptions listed in Table~\ref{ParamTable}.

\begin{table}[H]
\caption{Description of model parameters}
\small
\label{ParamTable}
\centering
\begin{tabular}{c l c}
\toprule
\textbf{Parameter} & \textbf{Description} & \textbf{Units} \\
\midrule
$\pi$      & Host recruitment rate & day$^{-1}$ \\

$\mu$      & Host natural mortality rate & day$^{-1}$ \\

$\delta$   & Host disease--induced mortality rate  & day$^{-1}$ \\

$\phi$     & Host recovery rate & day$^{-1}$ \\

$\alpha$   & Shedding rate of infected hosts into the environment & day$^{-1}$ \\

$\nu$      & Pathogen removal rate & day$^{-1}$ \\
\midrule
$\gamma$   & Relative host susceptibility & ---  \\

$\kappa$   & Environment--host contact rate & day$^{-1}$ \\

$c$        & Host--host contact rate & day$^{-1}$ \\
\midrule
$p$        & Host infection probability per environmental pathogen (in $f_W$) & ---  \\

$b$        & Host infection probability per infected host (in $f_I$) & ---  \\

$f_I$      & Dose--response for direct host--host transmission & --- \\

$\Lambda_I$ & Infectivity per unit dose (direct host-host transmission) & --- \\

$f_W$      & Dose--response for environment-to-host transmission & --- \\

$\Lambda_W$ & Infectivity per unit dose (environmental transmission) & --- \\

$\beta_W$  & Beta-Poisson shape parameter (environment) & --- \\

\bottomrule
\end{tabular}
\end{table}

\subsection{Dose-response functions}\label{sec:Dose_response}

In environmentally mediated transmission models, the dose-response function links exposure to infection risk and therefore directly shapes the force of infection.  Although different dose--response functional forms may coincide at very low doses and saturate at high doses, they can produce meaningfully different transmission dynamics when exposure occurs over intermediate dose ranges, where nonlinearity becomes important \cite{brouwer2017dose}.
We consider common functional forms for the dose--response functions $f_W(x)$ and
$f_I(x)$, where $x$ denotes the effective exposure dose determined by the
prevailing environmental pathogen load or infectious host prevalence,
respectively \cite{haas2015microbial,haas2014quantitative}. These functional forms are given by:

\begin{enumerate}[(i)]
\item Linear:
\begin{equation}\label{eq:linear}
f(x;\Lambda) = \Lambda x
\end{equation} 

\item Exponential:
\begin{equation}\label{eq:Exponential}
f(x;\Lambda) = 1 - e^{-\Lambda x}
\end{equation} 

\item Beta--Poisson (approximate):
\begin{equation}\label{eq:Beta_pois_approx}
f(x;\Lambda,\beta) = 1 - \left(1+\dfrac{x}{\beta}\right)^{-\Lambda \beta}
\end{equation} 

\end{enumerate}

\noindent Here, $\Lambda>0$ represents the per-unit-dose infectivity, a parameter governing how rapidly infection risk increases with small increases in pathogen dose at low dose, i.e., the low-dose slope since $\Lambda=f'(0)$. At low dose levels, all three functional forms therefore behave similarly, while nonlinear effects become important as the dose increases. In particular, the approximate Beta-Poisson form admits a single-hit interpretation under a constrained negative--binomial dose distribution with constant per-pathogen infectivity, subject to additional assumptions \cite{brouwer2017dose, haas2015microbial, haas2014quantitative}. Here, the parameter $\beta>0$ controls the curvature and saturation of the dose--response relationship.


\subsection{Incorporating periodicity}

Incorporating seasonal forcing is essential for modeling transmission dynamics of environmentally persistent pathogens, as both host susceptibility and environmental exposure often vary over time \cite{dowell2001seasonal,koelle2005pathogen}. Empirical studies have documented pronounced seasonal variation in infection severity and pathogen detectability in systems such as snake fungal disease \citep{mccoy2017environmental,mckenzie2019field}. 

To capture this, we model susceptibility ($\gamma$) and contact rate ($\kappa$ and $c$) parameters as positive, $\omega$-periodic functions of the form:
\begin{equation}\label{eq:generalseasonal}
\eta(t) \;=\; \bar{\eta}\Big(1 + A_{\eta}\,\sin\!\big(2\pi (t/\omega + \tau_{\eta})\big)\Big),
\quad 0 \le A_{\eta} \le 1,\;\; 0 \le \tau_{\eta} \le 1,
\end{equation}
where $\bar{\eta}$ denotes the mean value of the seasonal parameter $\eta(t)$, $A_{\eta}$ is the amplitude of periodic variation, and $\tau_{\eta}$ is the phase shift determining the timing of the seasonal peak within each period (See Appendix~\ref{appendix:tau} for details.). Positive values of $\tau_{\eta}$ advance the phase of ${\eta}(t)$, causing peaks to occur earlier in the seasonal cycle (Fig.~\ref{fig:seasonal_tau}).
Parameter definitions and admissible ranges for all terms appearing in the periodic
functions are summarized in Table~\ref{tab:periodic_symbols}. When two periodic
parameters  $i,j \in \{\gamma,\kappa,c\}$  share the same phase shift ($\tau_i=\tau_j$ for $i\neq j$), their seasonal dynamics are said to be \emph{synchronized}; otherwise, they are
\emph{asynchronous}.

\begin{table}[H]
\caption{Description of parameters for periodic functions}
\label{tab:periodic_symbols}
\centering
\begin{tabular}{cll}
\toprule
\textbf{Symbol} & \textbf{Description} & \textbf{Value} \\
\midrule
$A_i$ 
& Relative amplitude of seasonal variation for $i \in \{\kappa,c,\gamma\}$ 
& $0 \le A_i \le 1$ \\
\addlinespace

$\tau_i$ 
& Phase shift for $i \in \{\kappa,c,\gamma\}$
& $0 \le \tau_i \le 1$ \\
\addlinespace

$\omega$ 
& Seasonal period
& $365$ days\\

\addlinespace

$\bar{\gamma}$ 
& Average relative host susceptibility
& $1$ \\
\addlinespace
$\bar{\kappa}$ 
& Average environment-host contact rate
& $0.02$ \\
\addlinespace

$\bar{c}$ 
& Average host-host contact rate
& $0.02$ \\

\bottomrule
\end{tabular}
\end{table}

\subsection{Continuous-time Markov chain formulation}
To capture demographic and transmission variability that arises when the numbers of infectious hosts or environmental pathogen particles are small, we create a continuous-time Markov chain (CTMC) formulation of our system. For simplicity, we use the same notation for variables and parameters in the
CTMC model as in the ODE model, although all rates are now risks. The CTMC formulation extends the deterministic model in Eq.~\ref{eq:SIW} by treating all state variables as discrete random variables. In this setting, the state vector satisfies
\[
(S(t), I(t), W(t)) \in \mathbb{Z}_{\ge 0}^3, \qquad t \ge 0.
\]

The CTMC is a time–nonhomogeneous Markov process whose transition probabilities depend on the absolute time~$t$, through the seasonal functions $\gamma(t)$, $\kappa(t)$, and $c(t)$, and on the elapsed time between events. The disease-free state, characterized by
$(S,I,W)=(S^*,0,0)$ , is absorbing, since no infection or shedding events can occur
once all infected hosts and environmental pathogens are absent.

We denote the state of the system at time $t$ by $(s,i,w)$. For a sufficiently small $\Delta t > 0$, the infinitesimal transition probability to a new state $(j,k,\ell)$ at time $t+\Delta t$ is
\begin{multline}
p_{(s,i,w)\to(j,k,\ell)}(t,t+\Delta t)
= P\!\Big( (S,I,W)(t+\Delta t) = (j,k,\ell) 
\;\Bigm|\; (S,I,W)(t) = (s,i,w) \Big).
\end{multline}
All possible one-step transitions and their corresponding rates are listed in Table~\ref{tab:SIWctmc}. 

\begin{table}[H]
\caption{Transitions and rates for the CTMC model. For simplicity, we use the same notation for variables and parameters in the CTMC model as in the ODE model, although all rates are now risks}  \label{tab:SIWctmc}
\centering
\begin{tabular}{cllc}
\toprule
\textbf{Event} & \textbf{Description} & \textbf{State transition} & \textbf{Rate} \\
\midrule
1 & Recruitment of $S$      
  & $S \to S+1$             
  & $\pi(S+I)$ \\[2pt]

2 & Natural death of $S$    
  & $S \to S-1$             
  & $\mu S$ \\[2pt]

3 & Natural death of $I$    
  & $I \to I-1$             
  & $\mu I$ \\[2pt]

4 & Disease-induced death of $I$  
  & $I \to I-1$             
  & $\delta I$ \\[2pt]

5 & Recovery                 
  & $(I,S) \to (I-1, S+1)$  
  & $\phi I$ \\ \hline

6 & Infection via environment 
  & $(S,I)\to(S-1,I+1)$ 
  & $\gamma(t)\,\kappa(t)\, f_W(p W)\, S$ \\[2pt]

7 & Direct host--host infection 
  & $(S,I)\to(S-1,I+1)$ 
  & $\gamma(t)\, c(t)\, f_I(b I)\, S$ \\ \hline

8 & Shedding to environment  
  & $W \to W+1$             
  & $\alpha I$ \\[2pt]

9 & Environmental removal     
   & $W \to W-1$             
   & $\nu W$ \\ \hline

\end{tabular}

\end{table}

\section{Analytical results} \label{sec:analytic_results}

In this section we present analytical results for the model. We first characterize the disease-free equilibrium and compute the basic reproduction number in both constant and seasonal environments. A branching process approximation obtained
by linearizing the system about the disease-free equilibrium is used to derive
the probability of extinction when the outbreak is initiated by either
a single infectious host ($I=1$) or an equivalent amount of environmental pathogen ($W=1$). Finally,
we compare these analytical predictions with stochastic simulations of
the continuous-time Markov chain (CTMC) model and construct two-sided
Wilson confidence intervals for the estimated extinction probabilities.

\subsection{Disease-free equilibrium}

At the disease-free equilibrium (DFE), we have
\[
I^* =  W^* = 0.
\]
Thus, the susceptible equation reduces to
\[
\dot S = (\pi - \mu) S,
\]
so in the absence of infection, the host population follows exponential growth or decay depending on $\pi$ and $\mu$ ($\pi\neq\mu$). 
In our analysis, we assume demographic balance, $\pi = \mu$, so that the total population size remains constant and we write
\[
S^* = \bar N,
\]
where $\bar N$ denotes the resident host population size at the DFE. 
The DFE can then be written as
\[
(S^*, I^*, W^*) = (\bar N, 0, 0).
\]
In the CTMC formulation, the absorbing disease--free state is 
$$
\{(S,I,W): I=W=0\},
$$
with $S \ge 0$, which corresponds to pathogen extinction, since no further infection or shedding events can occur. However, the susceptible population may still vary through demographic events, so this set is not absorbing in the full process. The unique absorbing state for the CTMC formulation is $(0,0,0)$.

\subsection{Reproduction numbers in constant and periodic environments}\label{sec:R0_SFD}

Since the basic reproduction numbers associated with periodic and constant environments differ \cite{hridoy2025investigating, wang2008threshold}, 
we compute $\mathcal{R}_0$ for each case using the next-generation matrix approach, as described in Appendix~\ref{appendix:R0} including the infection and transition matrices $F(t)$ and $V$. In a constant environment, the periodic parameters are replaced by their average values,
i.e., $\kappa(t)\equiv\bar\kappa$, $c(t)\equiv\bar c$, and $\gamma(t)\equiv\bar\gamma$, and the basic reproduction number is given by

\begin{equation}\label{eq:R0_constant}
\mathcal R_0^{\mathrm{constant}}
=
\frac{1}{2}
\left[
\frac{\bar\gamma\bar c f_I'(0)b\bar N}{\mu+\delta+\phi}
+
\sqrt{
\left(
\frac{\bar\gamma\bar c f_I'(0)b\bar N}{\mu+\delta+\phi}
\right)^2
+
\frac{4\alpha\bar\gamma\bar\kappa f_W'(0)p\bar N}
{\nu(\mu+\delta+\phi)}
}
\right].
\end{equation}

For comparison with the seasonal environment, we also define the time-dependent instantaneous reproduction number
\begin{equation}\label{eq:R_inst_SIW}
\mathcal{R}^{\mathrm{inst}}(t_0)
=\rho\big(F(t_0)V^{-1}\big),\qquad t_0\in[0,\omega],
\end{equation}
which corresponds to freezing the seasonal parameters at time $t_0$ and computing
the basic reproduction number in the resulting constant environment. As the parameters are changing, the time-dependent instantaneous reproduction number is not a threshold condition for local asymptotic stability of the DFE.

In the seasonal environment, some terms are $\omega$--periodic due to seasonal variation in contact rates
and susceptibility, and the basic reproduction number, $\mathcal{R}_0^{\mathrm{per}}$, is therefore
computed using Floquet theory \cite{wang2008threshold, klausmeier2008floquet}.
An explicit closed form for $\mathcal{R}_0^{\mathrm{per}}$ is generally unavailable, so we compute
it numerically following the approach of Wang and Zhao \cite{wang2008threshold}, as described in Appendix \ref{appendix:R0}.

\subsection{Branching process approximation}\label{sec:BPA_SIW}

When infection is rare, such as during the early stages of an outbreak,
the stochastic dynamics of the system can be approximated by a branching
process obtained by linearizing the system around the DFE. This approximation provides analytical insight
into extinction probabilities and early outbreak behavior. 

For the branching process approximation (BPA) let 
\[
\mathbb{P}_{(i,w),(j,\ell)}(t_0,t)
=
P\!\Big(\big(I(t),W(t)\big)=(i,w)\,\big|\,\big(I(t_0),W(t_0)\big)=(j,\ell)\Big)
\] denote the transition probability.  
The initial states of the BPA correspond to: one infectious host, $(I=1,W=0)=\mathbf e_I$, or one unit of environmental pathogen, $(I=0,W=1)=\mathbf e_W.$
The absorbing state is disease extinction, $(I,W)=(0,0)$. Thus,
\[
\mathbb{P}_{\mathbf e_I}(t_0,t) := \mathbb{P}_{(1,0)}(t_0,t),
\qquad
\mathbb{P}_{\mathbf e_W}(t_0,t) := \mathbb{P}_{(0,1)}(t_0,t).
\]

Following the derivation detailed in Appendix~\ref{appendix:BPA}, the Backward Kolmogorov differential equations (BKDEs) for extinction probabilities, starting from a single I or W, take the general
form:

\[
-\,\frac{\partial \mathbb{P}_{\mathbf e_j}(t_0,t)}{\partial t_0}
=
\omega_j(t_0)\Big( g_j(t_0,\mathbb{P}_{\mathbf e_I},\mathbb{P}_{\mathbf e_W})
      -\mathbb{P}_{\mathbf e_j}(t_0,t) \Big),
\qquad j\in\{I,W\},
\]
where $g_j$ are the probability generating functions (pgf) of the linearized offspring distributions and $\omega_j(t_0)$ are the corresponding waiting-time functions. The pgf for the offspring of a single $I$ is
\begin{equation}\label{eq:pgf_I_SIW}
g_I\big(t_0,\mathbb{P}_{\mathbf e_I},\mathbb{P}_{\mathbf e_W}\big)
=
\frac{
      A_I(t_0)\,\mathbb{P}_{\mathbf e_I}^{\,2}
    + \alpha\,\mathbb{P}_{\mathbf e_I}\,\mathbb{P}_{\mathbf e_W}
    + D_I
     }
     {A_I(t_0)+\alpha+D_I}.
\end{equation}
and the pgf for the offspring of a single $W$ is
\begin{equation}\label{eq:pgf_W_SIW}
g_W\big(t_0,\mathbb{P}_{\mathbf e_I},\mathbb{P}_{\mathbf e_W}\big)
=
\frac{
      A_W(t_0)\,\mathbb{P}_{\mathbf e_I}\,\mathbb{P}_{\mathbf e_W}
    ,\mathbb{P}_{\mathbf e_W}^{\,2}
    + \nu
     }
     {A_W(t_0)+\nu}.
\end{equation}
where
\[
A_I(t_0) = \gamma(t_0)c(t_0)f_I'(0)\,b\,\bar N,
\qquad
D_I = \mu+\delta+\phi,
\qquad
A_W(t_0) = \gamma(t_0)\kappa(t_0)f_W'(0)\,p\,\bar N.
\]

The corresponding waiting–time functions are given by

\[
\omega_I(t_0)=A_I(t_0)+\alpha + D_I,
\qquad
\omega_W(t_0)=A_W(t_0) + \nu.
\]

\noindent The BPA allows us to quantify several outcome measures of early invasion and extinction dynamics in a seasonal environment, including:

\begin{enumerate}[(i)]
\item Probability of extinction, $\mathbb{P}_{\mathrm{ext}}$:
The extinction probability denotes the probability that the infection ultimately dies out when the outbreak is initiated at time $t_0$. Since the model parameters vary periodically over the seasonal cycle, it depends on the introduction time $t_0$ as well as on whether the outbreak is initiated by a single infectious host or a single environmental pathogen. We therefore define
\begin{equation}\label{eq:pext}
\begin{aligned}
\mathbb{P}_{\mathrm{ext}}^{j}(t_0), \quad j \in \{I,W\}.
\end{aligned}
\end{equation} 
Since closed-form solutions are generally unavailable, these extinction probabilities are obtained numerically by solving the BKDEs \eqref{eq:BKDE_I_single}--\eqref{eq:BKDE_W_single} over one period.

\item Average extinction probability, $\langle \mathbb{P}_{\mathrm{ext}} \rangle$: 

The average values of the probabilities of disease extinction, $\mathbb{P}_{\mathrm{ext}}^{j}(t_0)$, for initial conditions corresponding to an infected host or pathogen, $j = I, W$, are given by
\begin{equation}\label{eq:avgpext}
\begin{aligned}
\langle \mathbb{P}_{\mathrm{ext}}^{j} \rangle
=
\frac{1}{\omega}
\int_{0}^{\omega}
\mathbb{P}_{\mathrm{ext}}^{j}(t_0)\,dt_0.
\end{aligned}
\end{equation}

\item Phase shift of extinction extrema, $\theta$: 

To quantify the temporal offset between the extrema of the seasonal parameter and the extinction probability $\mathbb{P}_{\mathrm{ext}}^{j}(t_0)$, where $j\in\{I,W\}$, we define the extrema shifts as
\begin{equation}\label{eq:extremashift}
\theta_{\mathrm{min}}^{j}=t_{\mathrm{min}}^{j}-\widehat{t}_{\mathrm{max}},
\qquad
\theta_{\mathrm{max}}^{j}=t_{\mathrm{max}}^{j}-\widehat{t}_{\mathrm{min}}.
\end{equation}
Here, $t_{\mathrm{min}}^{j}$ and $t_{\mathrm{max}}^{j}$ denote the times at which $\mathbb{P}_{\mathrm{ext}}^{j}(t_0)$ attains its minimum and maximum over one seasonal period, respectively, while $\widehat{t}_{\mathrm{min}}$ and $\widehat{t}_{\mathrm{max}}$ denote the times at which the seasonal parameter attains its minimum and maximum. Negative values of $\theta_{\mathrm{min}}^{j}$ or $\theta_{\mathrm{max}}^{j}$ indicate that the corresponding extinction extremum occurs earlier in the seasonal cycle.

\end{enumerate}

\subsection{Estimation of extinction probabilities and confidence intervals from CTMC}\label{sec:CTMC_error}

A Monte Carlo simulation method is used to generate sample paths of the time-nonhomogeneous CTMC model. Simulations are performed using a sufficiently small time step $\Delta t$, chosen so that during each time step at most one event may occur, i.e., the sum of all transition probabilities satisfies $\sum_i r_i(t)\,\Delta t < 1$. At each step, a single event is selected according to the corresponding transition probabilities described in Table~\ref{tab:SIWctmc}, including the possibility that no event occurs. The accuracy of the Monte Carlo approximation is assessed by repeating simulations with progressively smaller time steps and verifying convergence of the resulting dynamics. While exact simulation methods exist for computing inter-event waiting times in time-dependent nonhomogeneous stochastic processes (see, e.g., \cite{lewis1979simulation, thanh2015simulation}), such methods are considerably more complex to implement for models with periodically varying transition rates. These exact approaches, commonly referred to as thinning or rejection-based algorithms, are therefore not pursued here. Instead, we adopt a Monte Carlo approach that has been shown to provide accurate approximations of time-nonhomogeneous CTMC dynamics when the time step is chosen sufficiently small (see, e.g., \cite{hridoy2025investigating, nipa2020disease}).

To quantify uncertainty in the CTMC estimates of extinction probabilities, we construct confidence intervals using the Wilson score method for binomial proportions \cite{wilson1927probable,brown2001interval}. For each initial time $t_0$, we run $n$ independent CTMC sample paths and define a disease extinction event as absorption at $(I,W)=(0,0)$ before the stopping criteria are met. Let $k(t_0)$ denote the number of extinct sample paths out of $n$, and let
\[
\widehat{\mathbb{P}}_{\mathrm{ext}}(t_0)=\frac{k(t_0)}{n}
\]
be the CTMC estimate of the extinction probability. To quantify sampling uncertainty in $\widehat{\mathbb{P}}_{\mathrm{ext}}(t_0)$, we compute two-sided $95\%$ Wilson confidence intervals for a binomial proportion. Using $z=1.96$, the Wilson confidence interval is given by
\begin{equation}\label{eq:Wilson}
\widehat{\mathbb{P}}_{\mathrm{ext}}^{\,\pm}(t_0)
=
\dfrac{
\widehat{\mathbb{P}}_{\mathrm{ext}}(t_0)
+\dfrac{z^{2}}{2n}
\ \pm\
z\sqrt{
\dfrac{\widehat{\mathbb{P}}_{\mathrm{ext}}(t_0)\bigl(1-\widehat{\mathbb{P}}_{\mathrm{ext}}(t_0)\bigr)}{n}
+\dfrac{z^{2}}{4n^{2}}
}
}{
1+\dfrac{z^{2}}{n}
}.
\end{equation}

\noindent
The Wilson interval provides improved finite-sample coverage relative to the normal approximation \cite{brown2001interval}, particularly when estimated probabilities are near $0$ or $1$, as commonly occurs in extinction calculations.


\subsection{Global sensitivity analysis}\label{sec:GSA}

We use global sensitivity analysis (GSA) to quantify how uncertainty in model parameters affects epidemiological quantities of interest (QoIs). In stochastic epidemic models variability arises from two distinct sources: uncertainty in the parameter values and intrinsic randomness. The latter comes from the stochastic nature of the Markov chain as repeated simulations of the CTMC model produce different outcomes even when parameters are fixed. Thus, we must distinguish between these two sources of variability and focus on the contribution from parameters rather than from intrinsic randomness.

Specifically, we perform a variance-based GSA using Sobol indices \cite{sobol2001global}. First-order indices quantify the direct contribution of individual parameters, whereas total-order indices capture their overall influence including interactions. Parameters with large total-order indices but relatively small first-order indices therefore indicate strong interaction effects. To assess the contribution of parameters to we estimate the first and total order Sobol indices as detailed in the Appendix~\ref{appendix:GSA}.

\section{Numerical results} \label{sec:num_results}

To show how environmental dose–response function, seasonal variability, and the timing of pathogen introduction can shape invasion, persistence, and extinction outcomes, we present numerical simulations of the model using snake fungal disease (SFD) as a case study. To begin, we describe the selection of model parameters from empirical studies and the literature. Then, we use simulations to show how the environmental dose–response function, seasonal variability, and the timing of pathogen introduction influence disease dynamics. We focus on the relative contributions of direct and environmental transmission and investigate phase shifts between seasonal transmission drivers and the resulting extinction probabilities.

\subsection{Model parameterization} \label{sec:param_SFD}

To illustrate the model
dynamics, the system is parameterized using biologically plausible
values motivated by the SFD literature, together with reasonable
assumptions where empirical estimates are unavailable. The parameter
values used in the simulations are listed in Table~\ref{SFDParamTable} and details on the derivation of these values are provided in
Appendix~\ref{appendix:param}.

The parameters $\gamma(t)$, $\kappa(t)$, and $c(t)$, which capture
seasonal variation in host susceptibility, environment-- host contact, and
host--host contact rates, respectively, are modeled as periodic
functions over biologically plausible ranges. In our numerical
simulations, we focus on the role of environmental transmission by
considering different functional forms for the environmental
dose--response function $f_W$
(Eqs.~\eqref{eq:linear}--\eqref{eq:Beta_pois_approx}), while the direct
transmission function $f_I$ is kept linear. Furthermore, to facilitate a direct comparison between transmission pathways, we set $\Lambda_I = \Lambda_W = 1$ and choose the average environment–host and host–host contact rates to be equal, $\bar{c} = \bar{\kappa}$. Therefore, the two transmission routes are comparable near the disease-free equilibrium, and differences in model behavior arise from the form of the dose–response functions and pathway-specific dynamics rather than from parameter scaling.

\begin{table}[H]
\caption{Parameter symbols, descriptions, baseline values, and  ranges used in simulations and global sensitivity analysis. References to literature derived values are provided, and parameters derived from empirical studies are described in more detail in Appendix \ref{appendix:param}. All range limits are from the literature unless otherwise indicated. \\$^*$Upper and lower limits of the range are determined by $0.25\times$ the baseline value. \\$^{\dagger}$Upper and lower limits of the range are determined by the limits of the seasonal amplitude.}
\footnotesize
\label{SFDParamTable}
\centering
\begin{tabular}{c l l l l c}
\toprule
\textbf{} & \textbf{Description} & \textbf{Value} &  \textbf{Range} & \textbf{Units} & \textbf{Source} \\
\midrule
$\pi$      & Host recruitment rate & $0.001$ &  $0.000258-0.001899$  & $\text{day}^{-1}$ & Assumed\\ 
$\mu$      & Natural mortality rate of hosts & $0.001$ & $0.000258-0.001899$  &$\text{day}^{-1}$ & \cite{weatherhead2012mortality} \\ 
$\delta$   & Disease–induced mortality rate of hosts & $0$ &  $0-0.0057$ & $\text{day}^{-1}$ & \cite{dillon2024effects,woah_sfd}  \\ 
$\phi$     & Host recovery rate  & $0.0602$  & $0.0051-0.07525^*$ & $\text{day}^{-1}$ & Experiment/ \cite{dillon2024effects} \\ 
$\alpha$   & Shedding rate of infected hosts & $0.03$ & $0.0225-0.0375 $ & $\text{day}^{-1}$ & \cite{mckenzie2020ophidiomycosis} \\ 
$\nu$          & Pathogen removal rate & $0.067$ & $0.05025-0.08375$ & $\text{day}^{-1}$ & Experiment/ \cite{campbell2021soil} \\[2pt] 
\hline
$\bar{\gamma}$          & Average relative host susceptibility & $1$ & $0^{\dagger}-2^{\dagger}$  &  -- & Varied \\ 
$\bar{\kappa}$   & Average environment contact rate for hosts & $0.02$ & $0^{\dagger}-0.04^{\dagger}$ & $\text{day}^{-1}$ &  Varied \\ 
$\bar{c}$ & Average host--host contact rate & $0.02$ & $0^{\dagger}-0.04^{\dagger}$ & $\text{day}^{-1}$ & Varied \\[2pt]
\hline
$p$        & Infection probability per particle for host (in $f_W$) & $0.01$ & $0.0075^*-0.0125^*$ & --  &  Assumed\\ 
$b$   & Linear scale factor (in $f_I$) & $0.01$ & $0.0075^*-0.0125^*$ & --  &  Assumed\\  
$\Lambda_I$    &  Infectivity per unit dose (direct host-host transmission) & $1$ &  $0.75^*-1.25^*$ & -- &  Assumed\\ 
$\Lambda_W$ & Infectivity per unit dose (environmental transmission) 
& $1$ &  $0.75^*-1.25^*$ & -- &  Assumed\\ 
$\beta_W$ & Beta-Poisson shape parameter (environment) 
& $0.40$ &  $0.30^*-0.50^*$ & -- &  Assumed\\
\bottomrule
\end{tabular}
\end{table}


\subsection{Effect of environmental dose–response function} \label{sec:NR_dose_response}

\begin{figure}[H]
\centering
\includegraphics[width=0.95\textwidth]{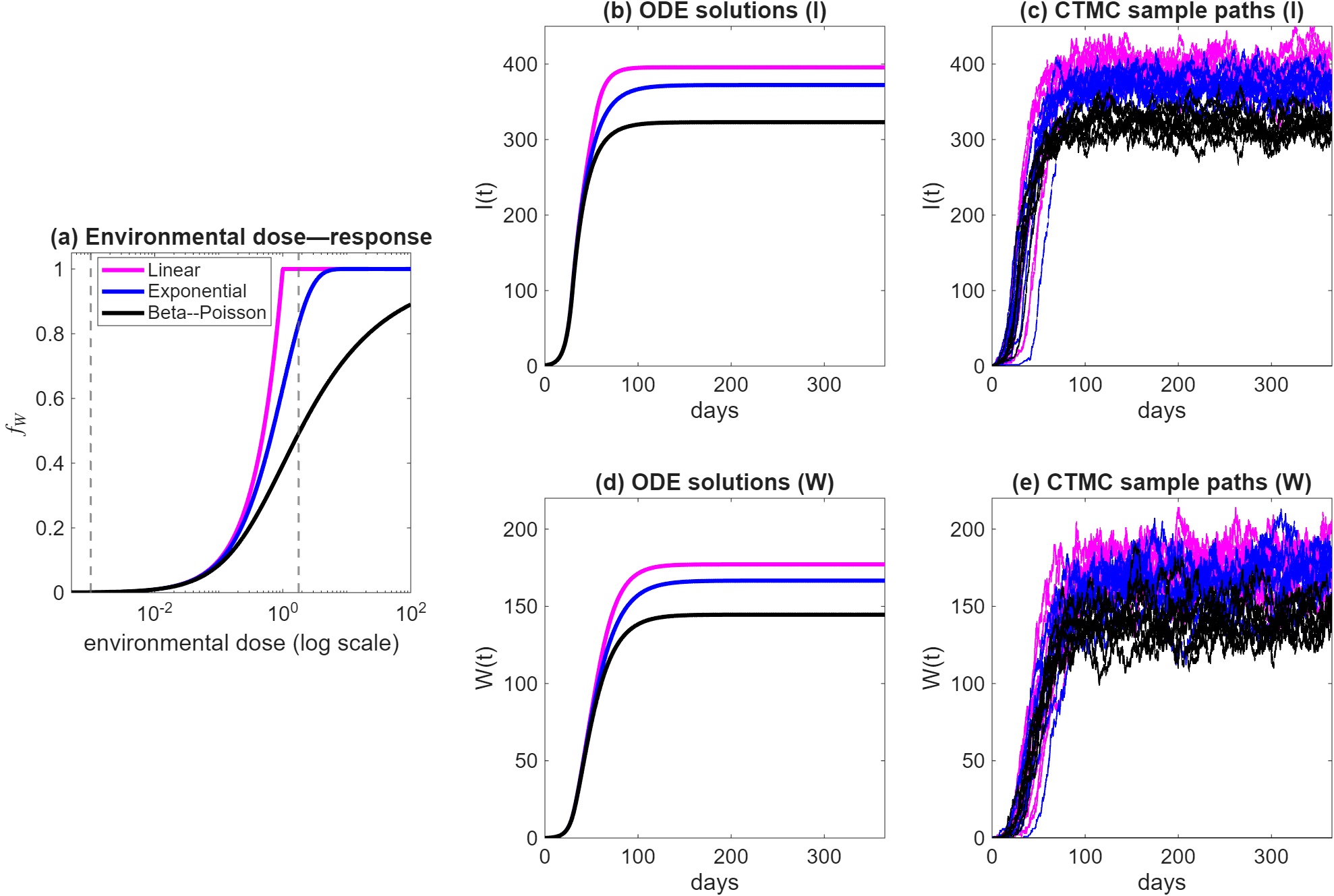} 
\caption{Model dynamics under different environmental dose--response
functions $f_W$, while direct transmission $f_I$ remains linear. Panel (a) shows the environmental dose--response functions $f_W(p_W W)$ evaluated over the range of environmental doses realized in the simulations (dashed lines). Panels (b) and (d) show the corresponding deterministic ODE solutions for infected hosts $I(t)$ and environmental pathogen load $W(t)$, respectively. Panels (c) and (e) show 10 CTMC sample paths for each dose--response case for $I(t)$ and $W(t)$, respectively. All simulations use the baseline parameter values in
Table~\ref{SFDParamTable} with initial conditions
$S(0)=1000$, $I(0)=1$, and $W(0)=0$.}\label{fig:dose_response_explore}
\end{figure}

We first examine how different environmental dose–response functions influence the system dynamics of both the deterministic ODE solutions and the stochastic CTMC sample paths (Fig.~\ref{fig:dose_response_explore}). Throughout, only the environmental dose–response function $f_W$ takes different functional forms, while direct transmission $f_I$ remains linear. 

As seen in Figure~\ref{fig:dose_response_explore}a, all three dose–response functions $f_W(p_W W)$ evaluated over the range of environmental doses share the same initial slope $f'_W(0)=\Lambda_W$, and therefore behave similarly at very low environmental doses. However, differences emerge at intermediate and higher doses. In particular, the linear and exponential dose--response functions increase more rapidly over the realized dose range than the Beta–Poisson form, which rises more gradually before approaching saturation. As a result, environmental transmission becomes effective earlier under the linear and exponential dose–response formulations, leading to faster epidemic growth and larger endemic levels in both the ODE solutions (Figs.~\ref{fig:dose_response_explore}b, \ref{fig:dose_response_explore}d) and the CTMC sample paths (Figs.~\ref{fig:dose_response_explore}c,~\ref{fig:dose_response_explore}e). Further comparisons of extinction outcomes across environmental dose–response formulations are provided in Table~\ref{Tab:ExtinctionBW} (Appendix~\ref{appendix:Extinction analysis}) for different dose–response functions and varying values of $\Lambda_W$.

\subsection{Effect of seasonality and introduction timing} \label{sec:NR_seasonality}

\begin{figure}[H]
\centering
\includegraphics[width=0.95\textwidth]{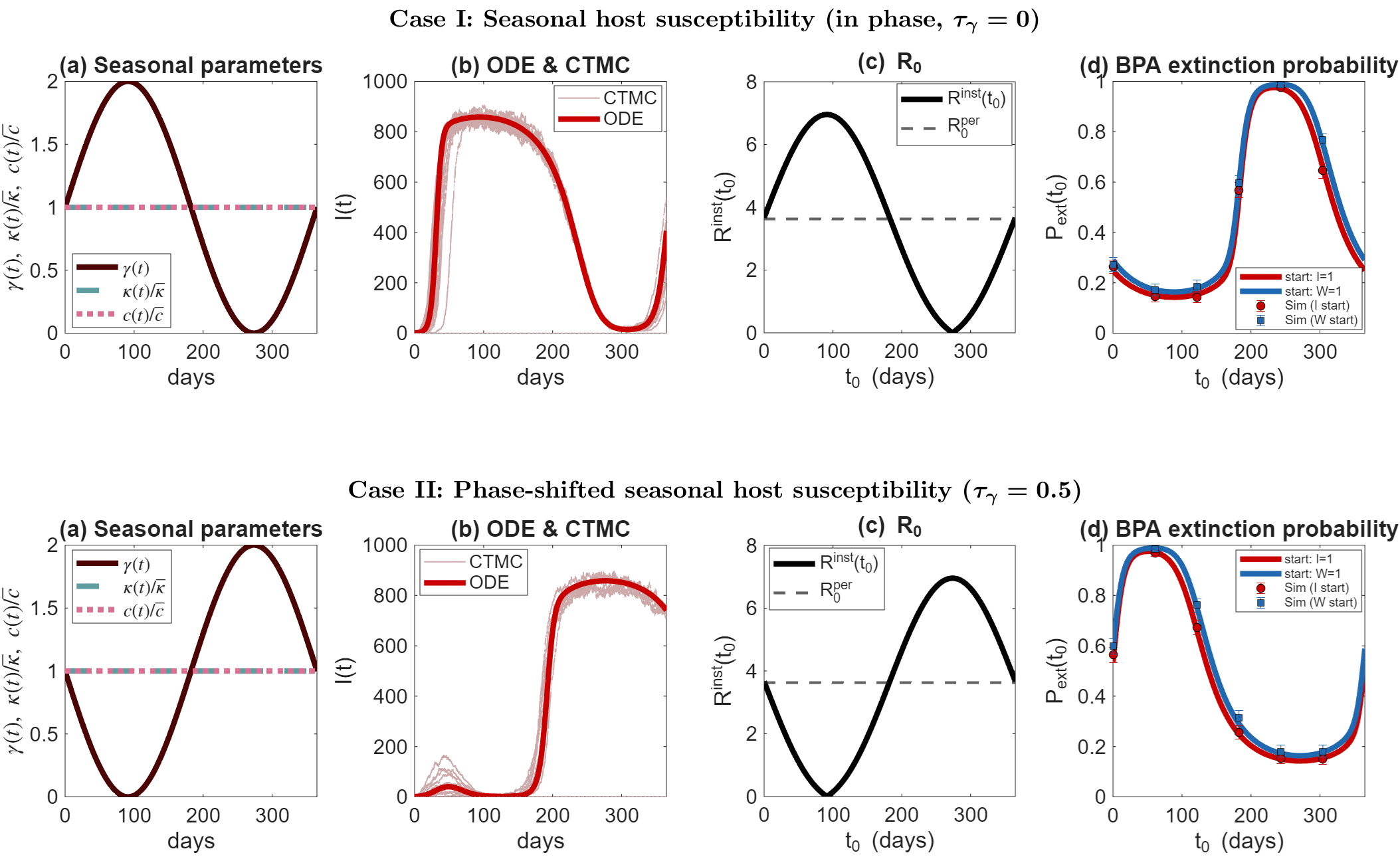} \caption{Seasonality in host susceptibility $\gamma(t)$, following the seasonal form in Eq.~\eqref{eq:gamma} with $A^{\gamma}=1$, drives disease dynamics and probability of extinction. Panel (a) shows seasonal variation in $\gamma(t)$. Panel (b) shows the ODE solution for $I(t)$ with 10 CTMC sample paths. Panel (c) shows instantaneous reproduction number $\mathcal{R}^{\mathrm{inst}}(t_0)$ and seasonal reproduction number $\mathcal{R}_0^{\mathrm{per}}=3.628$. Panel (d) shows probability of extinction from the BPA, starting from $I=1$ or $W=1$, compared with estimates from 1000 CTMC simulations (error bars show 95\% Wilson confidence intervals). Case I: $\tau_{\gamma}=0$; Case II: $\tau_{\gamma}=0.5$.}
\label{fig:only_gamma_per}
\end{figure}

We examine how seasonal variation influences infection prevalence, the instantaneous reproduction number $\mathcal{R}^{\mathrm{inst}}(t_0)$, the seasonal reproduction number $\mathcal{R}_0^{\mathrm{per}}$, and the seasonal extinction probabilities obtained from the BPA, $\mathbb{P}_{\mathrm{ext}}^{j}(t_0)$, $j \in \{I,W\}$. 
We first consider seasonal variation in $\gamma(t)$ while $\kappa(t)$ and $c(t)$ remain constant (Fig.~\ref{fig:only_gamma_per}) through two seasonal scenarios: Case I, where $\gamma(t)$ experiences no phase shift ($\tau_{\gamma}=0$), and Case II, where its seasonal cycle is shifted by half a period ($\tau_{\gamma}=0.5$). In both cases, the deterministic ODE solution and CTMC sample paths show similar epidemic timing, though stochastic trajectories fluctuate around the deterministic trend (Fig.~\ref{fig:only_gamma_per}b). BPA extinction probabilities agree closely with estimates from $1000$ CTMC simulations, with 95\% Wilson confidence intervals overlapping the branching-process predictions (Fig.~\ref{fig:only_gamma_per}d).

The extinction curves for introductions through infected hosts $\mathbb{P}_{\mathrm{ext}}^{I} (t_0)$ and through the environmental reservoir $\mathbb{P}_{\mathrm{ext}}^{W} (t_0)$ are broadly similar: introductions during low host susceptibility are much more like to go extinct, whereas persistence is likely when susceptibility increases. Therefore, seasonal variation in $\gamma(t)$ modulates both direct and environmental transmission pathways in a similar manner. However, introducing the phase shift substantially alters the timing of invasion risk as seen in Case II where the seasonal cycle begins with low host susceptibility, so early introductions are more likely to go extinct. 

This shift in invasion windows is also clearly reflected in the BPA extinction probabilities. Furthermore, the BPA extinction probabilities exhibit clear seasonal structure that closely follows the patterns predicted by the instantaneous reproduction number (Figs.~\ref{fig:only_gamma_per}c, ~\ref{fig:only_gamma_per}d): periods with larger $\mathcal{R}^{\mathrm{inst}}(t_0)$ correspond to lower extinction probabilities, whereas periods with smaller $\mathcal{R}^{\mathrm{inst}}(t_0)$ correspond to higher extinction probabilities.

\begin{figure}[H]
\centering
\includegraphics[width=0.95\textwidth]{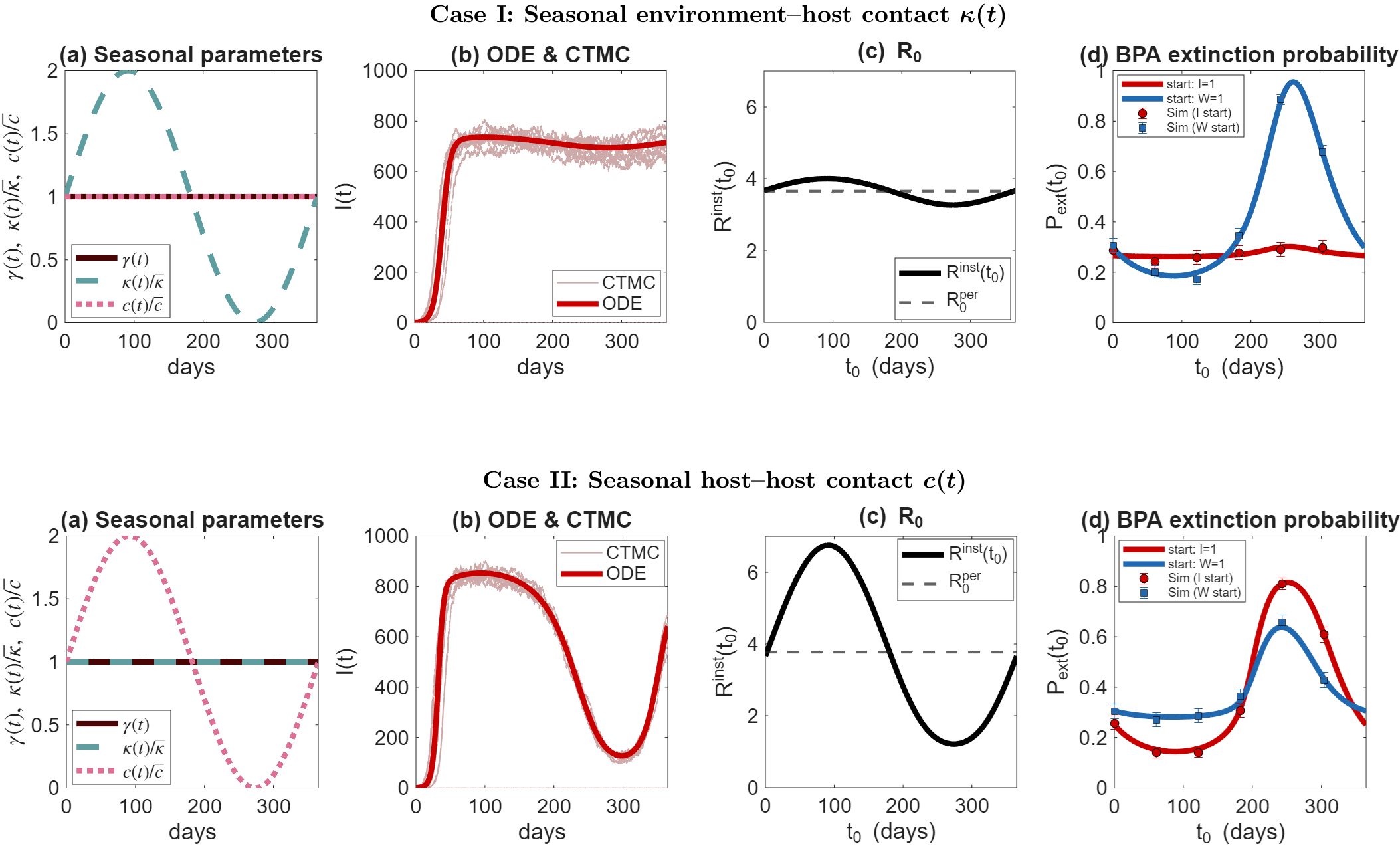} \caption{Seasonality in contact parameters differentially alters disease dynamics and probability of extinction. Case I: Only environment--host contact rate $\kappa(t)$ is periodic with $A^{\kappa}=1$, $\tau_{\kappa}=0$, and $\mathcal{R}_0^{\mathrm{per}}=3.653$; Case II: Only host--host contact rate $c(t)$ is periodic with $A^{c}=1$, $\tau_{c}=0$, and $\mathcal{R}_0^{\mathrm{per}}=3.777$.}
\label{fig:kappa_c_per}
\end{figure}

We next consider seasonal variation in the environment–host contact rate $\kappa(t)$ (Case I) and host–host contact rate $c(t)$ (Case II) while host susceptibility $\gamma(t)$ remains constant (Fig.~\ref{fig:kappa_c_per}). When only $\kappa(t)$ varies seasonally (Case I), infection prevalence exhibits relatively weak seasonal variation (Fig.~\ref{fig:kappa_c_per}b, Case I), and the instantaneous reproduction number $\mathcal{R}^{\mathrm{inst}}(t_0)$ changes only modestly and remains close to the periodic threshold $\mathcal{R}_0^{\mathrm{per}}$ (Fig.~\ref{fig:kappa_c_per}c, Case I).  However, we observe that extinction risk depends strongly on the route of introduction such that seasonal variation in $\kappa(t)$ substantially alters extinction probabilities for introductions through the environmental reservoir $\mathbb{P}_{\mathrm{ext}}^{W}(t_0)$ while having comparatively little effect on introductions through infected hosts $\mathbb{P}_{\mathrm{ext}}^{I}(t_0)$ (Fig.~\ref{fig:kappa_c_per}d, Case I). In contrast, seasonal variation in $c(t)$ produces much larger fluctuations in $\mathcal{R}^{\mathrm{inst}}(t_0)$ and stronger temporal changes in epidemic prevalence (Figs.~\ref{fig:kappa_c_per}b and ~\ref{fig:kappa_c_per}c, Case II). The associated extinction probabilities vary substantially over the seasonal cycle for both introduction pathways, with especially pronounced changes for host introductions (Fig.~\ref{fig:kappa_c_per}d, Case II). Thus, compared with seasonal variation in $\kappa(t)$, variation in $c(t)$ influences invasion dynamics across both introduction pathways, whereas seasonal variation in $\kappa(t)$ primarily affects invasion following environmental introductions.

\begin{figure}[H]
\centering
\includegraphics[width=0.95\textwidth]{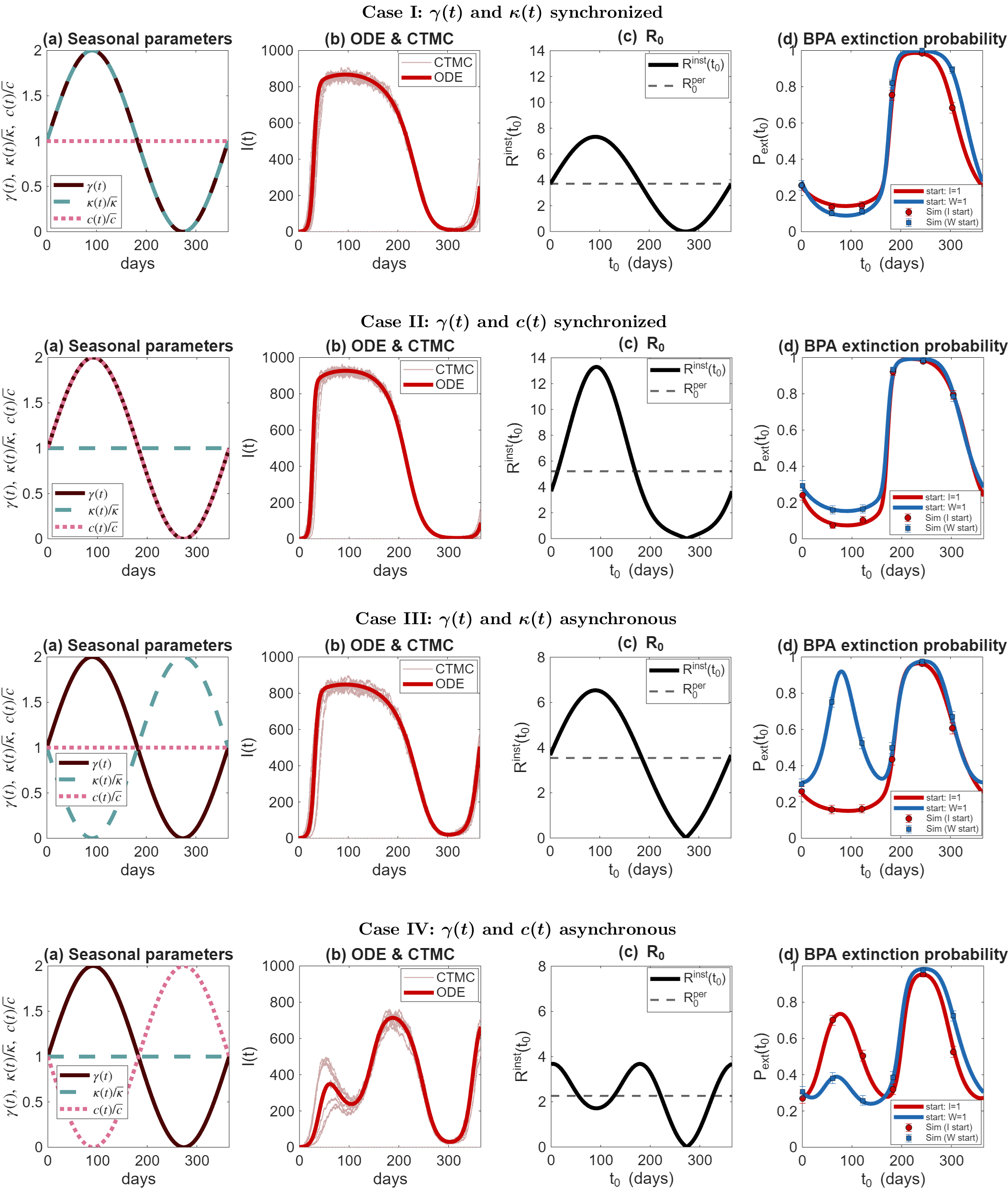} \caption{Synchrony in the seasonality of host susceptibility and contact parameters combines to affect disease dynamics and probability of extinction. Case I: Both $\gamma(t)$ and $\kappa(t)$ are periodic with $A^{\gamma}=1=A^{\kappa}$, $\tau_{\gamma}=0= \tau_{\kappa}$ ($\mathcal{R}_0^{\mathrm{per}}=3.699$); Case II: Both $\gamma(t)$ and $c(t)$ are periodic with $A^{\gamma}=1=A^{c}$, $\tau_{\gamma}=0= \tau_{c}$ ($\mathcal{R}_0^{\mathrm{per}}=5.213$); Case III: Both $\gamma(t)$ and $\kappa(t)$ are periodic with $A^{\gamma}=1=A^{\kappa}$, $\tau_{\gamma}=0, \tau_{\kappa}=0.5$ ($\mathcal{R}_0^{\mathrm{per}}=3.541$); Case IV: Both $\gamma(t)$ and $c(t)$ are periodic with $A^{\gamma}=1=A^{c}$, $\tau_{\gamma}=0, \tau_{c}=0.5$ ($\mathcal{R}_0^{\mathrm{per}}=2.262$).}
\label{fig:sync_async}
\end{figure}

To examine the effect of synchrony and asynchrony between seasonal drivers, we consider scenarios in which host susceptibility $\gamma(t)$ varies together with either the environment–host contact rate $\kappa(t)$ or the host–host contact rate $c(t)$ (Fig.~\ref{fig:sync_async}). Specifically, we compare cases in which the seasonal drivers are synchronized (Cases I and II) with those in which they are asynchronous through the phase shift $\tau$ (Cases III and IV). 

When the seasonal drivers are synchronized, fluctuations in the instantaneous reproduction number $\mathcal{R}_0^{\mathrm{inst}}(t)$ and the extinction probabilities become more pronounced, accentuating the patterns observed when each parameter varies individually (Figs.~\ref{fig:sync_async}c and ~\ref{fig:sync_async}d, Cases I and II). In particular, synchronizing $\gamma(t)$ with direct contact $c(t)$ (Case II) produces the strongest seasonal fluctuation, yielding the largest annual variation in $\mathcal{R}_0^{\mathrm{inst}}(t)$ and the highest periodic reproduction number $\mathcal{R}_0^{\mathrm{per}}$ among all four cases. In both synchronized scenarios, periods of high susceptibility coincide with high transmission, creating favorable invasion windows and correspondingly lower extinction probabilities. Furthermore, the effect of synchrony depends on the route of introduction. When $\gamma(t)$ is synchronized with the seasonal transmission pathway, introductions through that same pathway have lower extinction risk during favorable periods (Fig.~\ref{fig:sync_async}d, Cases I and II).

In contrast, when the seasonal drivers for susceptibility and contact rate are asynchronous, the relative timing between host susceptibility and the seasonal transmission pathway as well as the introduction route play a key role. 
In particular, when host susceptibility $\gamma(t)$ and environment-to-host contact rate $\kappa(t)$ are out of phase, 
the misalignment enhances the windows of extinction risk for environmental introductions, with $\mathbb{P}_{\mathrm{ext}}^{W}(t_0)$ showing marked peaks when both susceptibility and environment-to-host contact rate are low (Fig.~\ref{fig:sync_async}d, Case III. 
In comparison, the extinction probabilities for host introductions are highly sensitive to host susceptibility as host-to-host contact $c(t)$ remains constant with limit impact from environment-to-host contact variability.
When $\gamma(t)$ and host-to-host contact rate $c(t)$ are out of phase, both introduction pathways display similar seasonal timing in extinction risk, although introductions through infected hosts generally exhibit higher extinction probabilities during unfavorable windows (Fig.~\ref{fig:sync_async}d, Case IV). 
These patterns suggest that variability in host-to-host transmission has a larger impact on epidemic dynamics than does environment-to-host transmission.

\begin{figure}[H]
\centering
\includegraphics[width=0.95\textwidth]{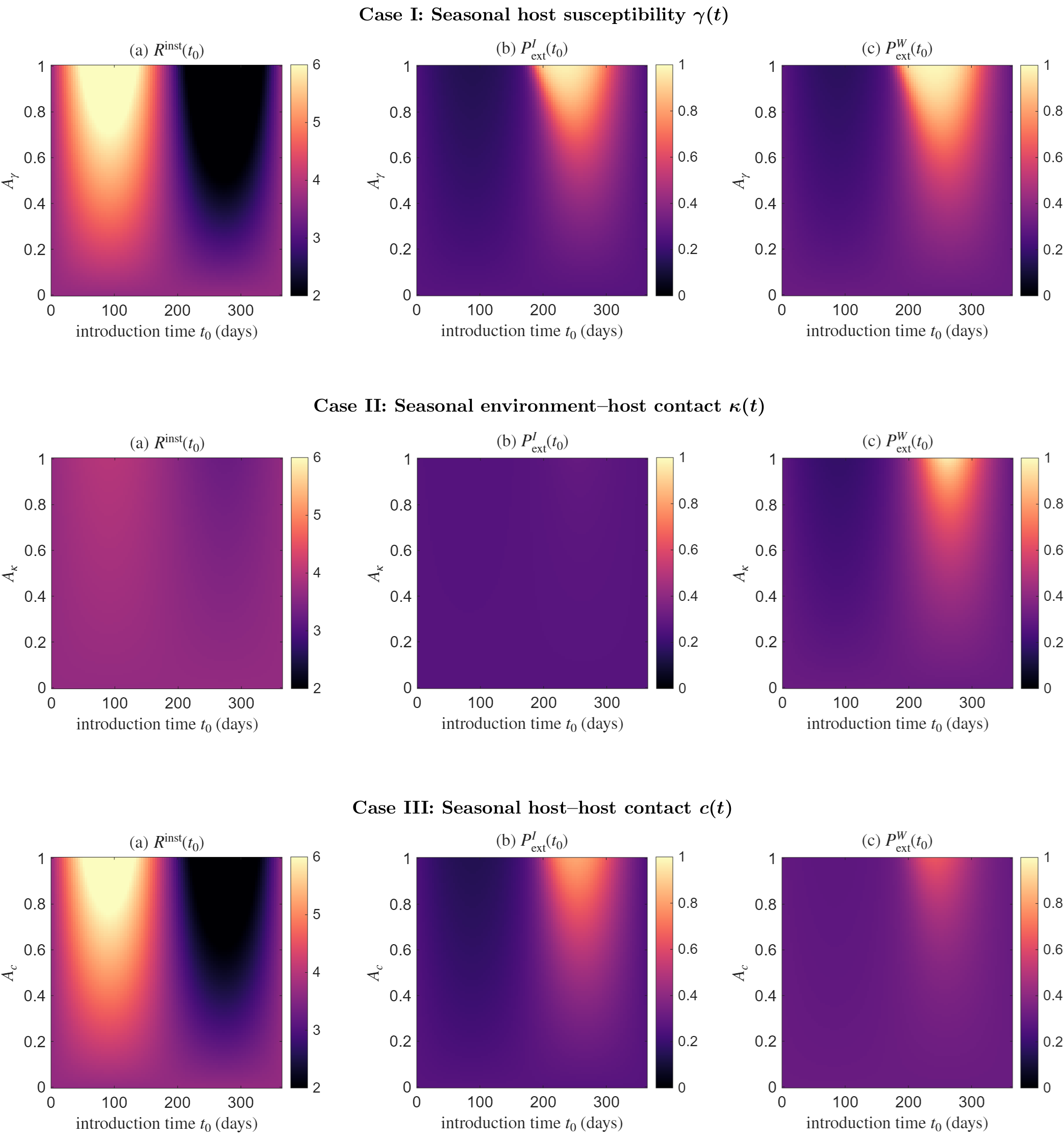} \caption{Varying the seasonal amplitude $A_i$ ($i \in \{\gamma,\kappa,c\}$) and the time of introduction $t_0$ of either a single infectious host or a single environmental pathogen. Columns show (a) the instantaneous reproduction number $\mathcal{R}^{\mathrm{inst}}(t_0)$, (b) $\mathbb{P}_{\mathrm{ext}}^{I}(t_0)$ for introductions initiated by a single infected host, and 
(c) $\mathbb{P}_{\mathrm{ext}}^{W}(t_0)$ for introductions initiated by a single environmental pathogen from BPA. Case I: seasonal variation in $\gamma(t)$, 
Case II: seasonal variation in $\kappa(t)$, and 
Case III: seasonal variation in $c(t)$.}
\label{fig:heatmap_A_intro_all}
\end{figure}

Figure~\ref{fig:heatmap_A_intro_all} summarizes the effects of varying the seasonal amplitude $A_i$ ($i \in \{\gamma,\kappa,c\}$) and the time of introduction $t_0$ of either a single infectious host or a single environmental pathogen. As the seasonal amplitude approaches zero ($A_i \to 0$), all dynamics become nearly uniform, consistent with the constant-environment case.
When either $\gamma(t)$ or $c(t)$ vary seasonally, the instantaneous reproduction number $\mathcal{R}^{\mathrm{inst}}(t_0)$ shows pronounced dependence on both $t_0$ and amplitude, with stronger seasonal variation producing larger contrasts between favorable and unfavorable introduction windows (Fig.~\ref{fig:heatmap_A_intro_all}a, Cases I and III). In contrast, when only $\kappa(t)$ varies (Case II), changes in $\mathcal{R}^{\mathrm{inst}}(t_0)$ are comparatively modest across both amplitude and introduction time, indicating a weaker effect of environmental contact seasonality on overall transmission potential, consistent with earlier results.
These differences are reflected in the extinction probabilities. In general, introductions occurring during periods with larger $\mathcal{R}^{\mathrm{inst}}(t_0)$ correspond to lower extinction probabilities, whereas introductions during less favorable periods are more likely to go extinct. Seasonal variation in $\gamma(t)$ and $c(t)$ therefore generates pronounced regions of low and high extinction risk, especially as amplitude increases (Figs.~\ref{fig:heatmap_A_intro_all}b and ~\ref{fig:heatmap_A_intro_all}c, Cases I and III). In contrast, seasonal variation in $\kappa(t)$ has little effect on extinction probabilities for host introductions, $\mathbb{P}_{\mathrm{ext}}^{I}(t_0)$, but substantially alters extinction risk for environmental introductions, $\mathbb{P}_{\mathrm{ext}}^{W}(t_0)$, since this pathway depends directly on environment--host contact. In contrast, when only $c(t)$ varies, extinction risk for environmental introductions also changes seasonally, but less strongly than for host introductions because the effect is mediated indirectly through subsequent host transmission. A similar analysis varying introduction time and phase shift $\tau_i$ is found in Appendix \ref{appendix:intro_shift}.

\subsection{Effect of seasonal phase shifts} \label{sec:NR_seasonality}

\begin{figure}[H]
\centering
\includegraphics[width=0.75\textwidth]{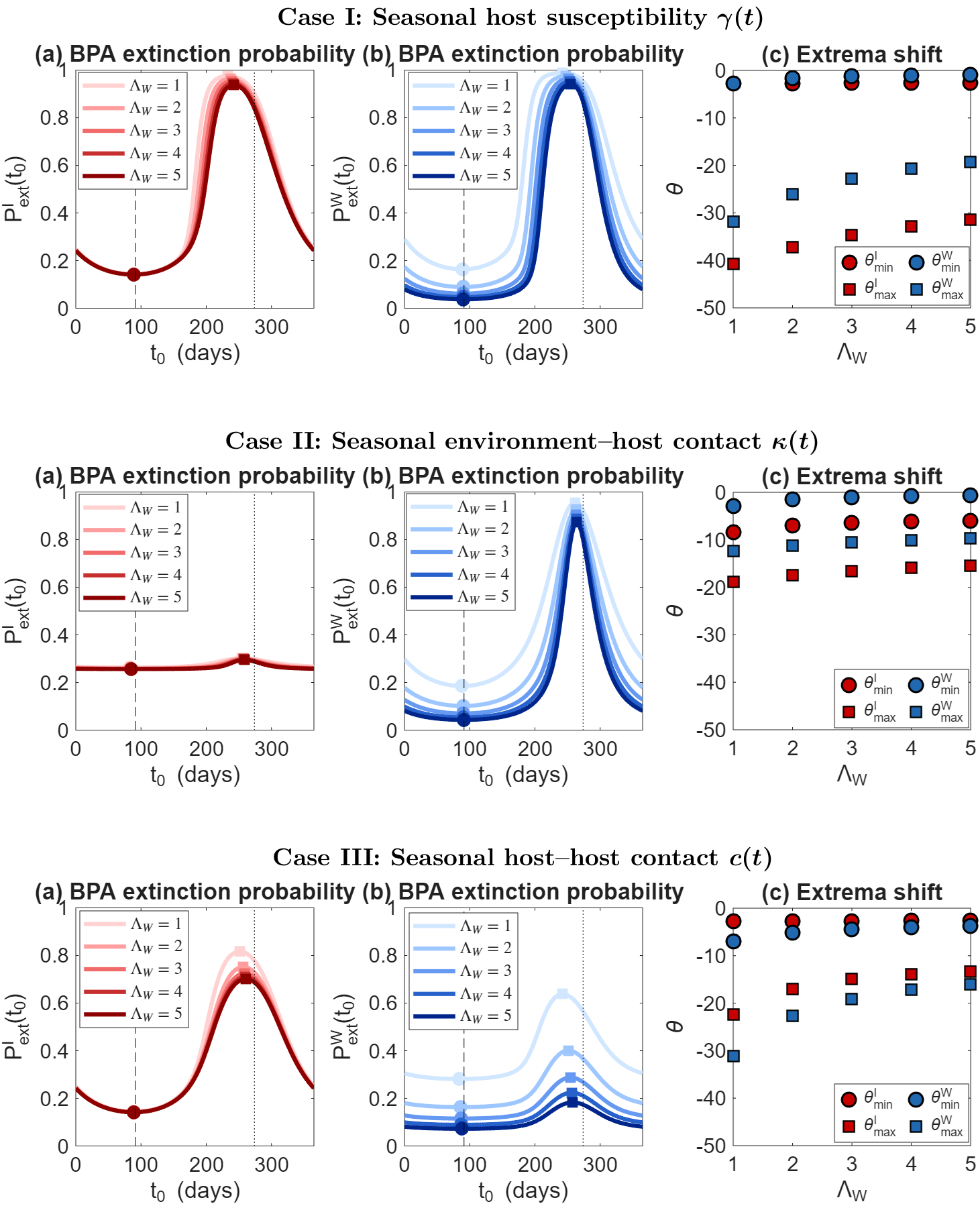} 
\caption{Extrema shift $\theta$ with variation in the environmental initial low-dose slope $f_W'(0)=\Lambda_W$. Panels (a) and (b) show the branching-process extinction probabilities $\mathbb{P}_{\mathrm{ext}}^{I}(t_0)$ and $\mathbb{P}_{\mathrm{ext}}^{W}(t_0)$, respectively, for varying environmental initial low-dose slope $f_W'(0)=\Lambda_W$. Vertical dashed and dotted black lines indicate the seasonal maximum and minimum of the periodic parameter, respectively, in each case. Panel (c) shows the corresponding extrema shifts $\theta_{\min}^{j}$ and $\theta_{\max}^{j}$, $j\in\{I,W\}$, defined as in Eq.~\eqref{eq:extremashift}. Case I: only $\gamma(t)$ is periodic; Case II: only $\kappa(t)$ is periodic; Case III: only $c(t)$ is periodic.}
\label{fig:P_ext_diff_LambdaW}
\end{figure}

Across all cases, extrema in extinction probability occur slightly before the corresponding extrema of $\mathcal{R}^{\mathrm{inst}}(t_0)$, indicating a temporal offset between transmission potential and the associated invasion window. To quantify this effect, we examine the extrema shift $\theta$, defined in Eq.~\eqref{eq:extremashift} as the difference between the time at which the extinction probability attains a local extremum and the corresponding extremum of the seasonal driver. We vary the environmental initial low-dose slope $f_W'(0)=\Lambda_W$, 
which governs how rapidly infection risk increases with small increases in pathogen dose near zero, 
and compute extinction probabilities from BPA, 
starting from either a single infected host or environmental pathogen under three seasonal scenarios in which only one parameter varies periodically at a time (Fig.~\ref{fig:P_ext_diff_LambdaW}).

In all cases, the timing of peak extinction probability does not coincide exactly with the extrema of the seasonal parameter, reflecting the interaction between seasonal transmission and infection dynamics. Extinction extrema generally occur slightly before the corresponding extrema of the seasonal driver, as indicated by negative shifts (Fig.~\ref{fig:P_ext_diff_LambdaW}c).

Extinction dynamics more closely track seasonal variation along the transmission pathway that is directly affected by the seasonal parameter, while the alternative pathway can exhibit a larger temporal lag. When $\gamma(t)$ varies seasonally, shifts are observed for both introduction pathways, consistent with susceptibility modulating both direct and environmental transmission (Fig.~\ref{fig:P_ext_diff_LambdaW}, Case I). When $\kappa(t)$ varies seasonally, the shift is more pronounced for host introductions, $\mathbb{P}_{\mathrm{ext}}^{I}(t_0)$, than for environmental introductions, $\mathbb{P}_{\mathrm{ext}}^{W}(t_0)$ (Fig.~\ref{fig:P_ext_diff_LambdaW}, Case II). In contrast, when $c(t)$ varies seasonally, the larger shift occurs for environmental introductions, particularly for the maxima of $\mathbb{P}_{\mathrm{ext}}^{W}(t_0)$ (Fig.~\ref{fig:P_ext_diff_LambdaW}, Case III).

Stronger transmission intensity diminishes the temporal lag between the seasonal driver and extinction dynamics. In particular, increasing $\Lambda_W$ reducing the extrema shift toward zero, aligning them more closely with those of the seasonal parameter. Since the BPA depends only on the low-dose slope $f_W'(0)=\Lambda_W$, increasing $\Lambda_W$ increases effective transmission intensity and therefore decreases extinction probabilities, with the stronger effect observed for environmental introductions. Furthermore, the shifts associated with the maxima, $\theta_{\max}^{j}$, are consistently smaller in magnitude than those associated with the minima, $\theta_{\min}^{j}$.

\begin{figure}[H]
\centering
\includegraphics[width=\textwidth]{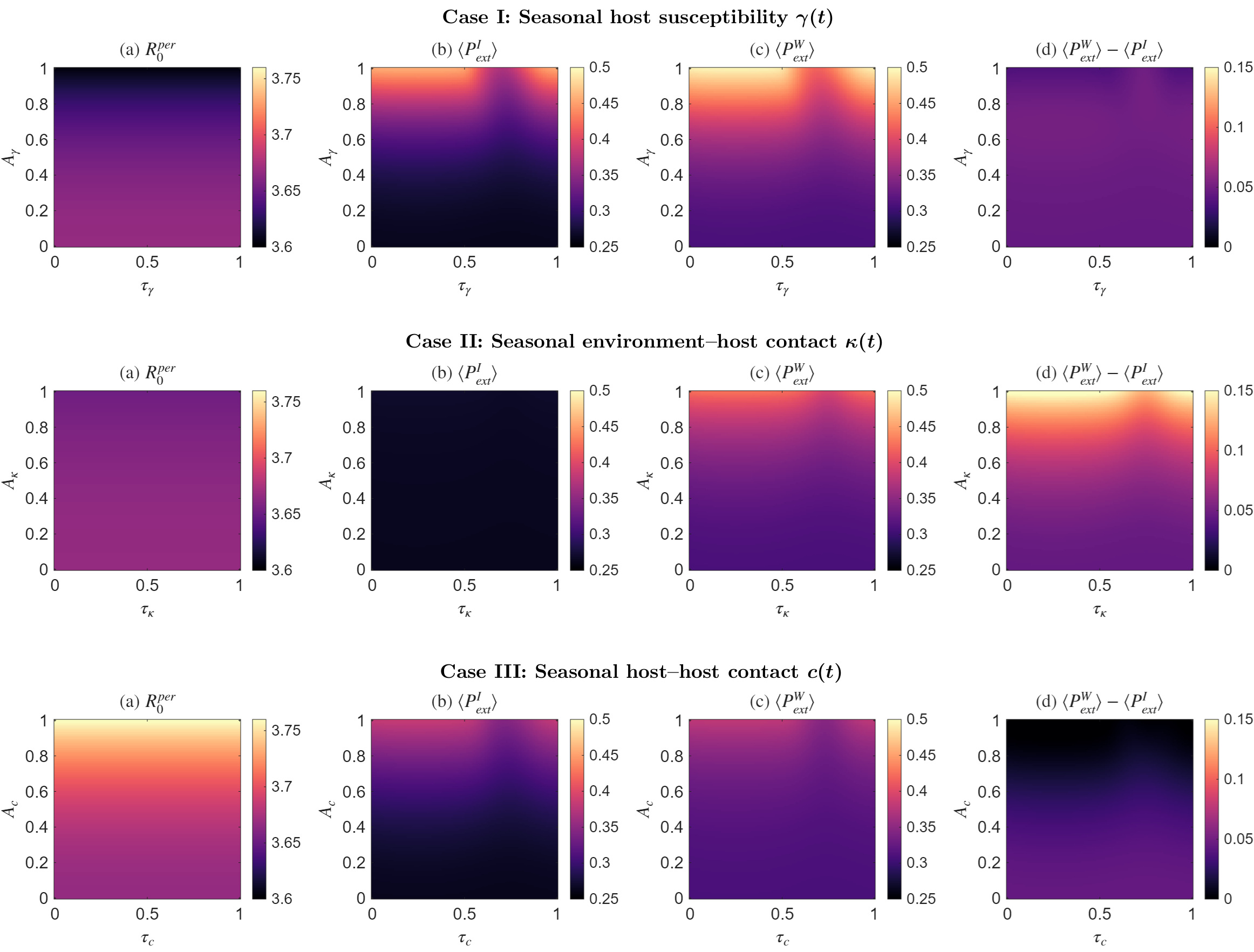} \caption{Varying the seasonal amplitude $A_i$ ($i \in \{\gamma,\kappa,c\}$) and the phase shift $\tau_i$ with the introduction of either a single infectious host or a single environmental pathogen.
Columns show (a) the seasonal reproduction number $\mathcal{R}_0^{\mathrm{per}}$, 
(b) the average extinction probability $\langle \mathbb{P}_{\mathrm{ext}}^{I} \rangle$ for introductions initiated by a single infected host, 
(c) the average extinction probability $\langle \mathbb{P}_{\mathrm{ext}}^{W} \rangle$ for introductions initiated by a single environmental pathogen, 
and (d) the difference $\langle \mathbb{P}_{\mathrm{ext}}^{I} \rangle - \langle \mathbb{P}_{\mathrm{ext}}^{W} \rangle$. 
Case I: seasonal variation in host susceptibility $\gamma(t)$, 
Case II: seasonal variation in environmental contact $\kappa(t)$, and 
Case III: seasonal variation in host--host contact $c(t)$.}
\label{fig:heatmap_A_tau_all}
\end{figure}

We further examine the combined effects of seasonal amplitude and phase shift by varying the amplitude $A_i$ and phase $\tau_i$ of the seasonal parameters ($i \in \{\gamma,\kappa,c\}$) and compute the seasonal reproduction number $\mathcal{R}_0^{\mathrm{per}}$, the average BPA extinction probabilities $\langle \mathbb{P}_{\mathrm{ext}}^{I} \rangle$ and $\langle \mathbb{P}_{\mathrm{ext}}^{W} \rangle$, and their difference (Fig.~\ref{fig:heatmap_A_tau_all}).

In all three cases, $\mathcal{R}_0^{\mathrm{per}}$ varies only with the amplitude $A_i$ of the seasonal parameter. The effect of increasing seasonal amplitude $A_i$ differs across the three cases. For seasonal $\gamma(t)$ or $\kappa(t)$ (Cases I and II), increasing the amplitude reduces $\mathcal{R}_0^{\mathrm{per}}$. 
In contrast, for seasonal $c(t)$ (Case III), larger amplitudes lead to an increase in $\mathcal{R}_0^{\mathrm{per}}$. This difference highlights how seasonal variability affects different transmission routes: stronger fluctuations in host susceptibility and environmental contact tends to introduce longer periods of reduced transmission, which lowers overall invasion potential, whereas stronger fluctuations in direct host-to-host contact can create short periods of particularly favorable transmission that increase pathogen growth over the seasonal cycle.
Across all three cases, increasing $A_i$ generally leads to higher average extinction probabilities with distinct windows created by $\tau_i$ during the year where pathogen introduction is either more or less likely to result in successful invasion (Figs.~\ref{fig:heatmap_A_tau_all}b and \ref{fig:heatmap_A_tau_all}c).

$\mathcal{R}_0^{\mathrm{per}}$ shows no dependence on the phase shift $\tau_i$, because changing $\tau_i$ only shifts the seasonal parameter in time without altering its overall shape across a full period. Since $\mathcal{R}_0^{\mathrm{per}}$ is computed using Floquet theory, it depends only on the magnitude of seasonal variation rather than the timing of seasonal peaks. 
In contrast, the BPA extinction probabilities vary substantially with $\tau_i$. This reflects the dependence of invasion success on the timing of pathogen introduction relative to seasonal conditions.
Importantly, $\tau_i$ influences extinction probabilities not only through the timing of introduction relative to seasonal peaks, but also through the trajectory of transmission conditions immediately following introduction (Fig.~\ref{fig:seasonal_tau}). In particular, whether the seasonal parameter is increasing or decreasing at the time of introduction plays an important role in determining invasion success. For example, when $\tau_i = 0.25$, the seasonal curve begins near its maximum at $t=0$, providing favorable transmission conditions immediately after introduction. More generally, for $0 < \tau_i < 0.25$ the seasonal parameter begins at relatively high values and continues increasing toward the seasonal peak, which corresponds to lower extinction probabilities (Figs.~\ref{fig:heatmap_A_tau_all}b and \ref{fig:heatmap_A_tau_all}c). In contrast, for $0.25 < \tau_i < 0.5$, the system begins near intermediate transmission levels but immediately enters a declining phase following the seasonal peak. Thus, extinction risk is governed not only by the instantaneous transmission level at the time of introduction, but also by whether transmission conditions improve or deteriorate immediately afterward. This same qualitative structure appears in all three seasonal scenarios, although the magnitude of variation differs by parameter.

Throughout, average extinction probabilities associated with environmental introduction remain consistently higher than those associated with host introduction, i.e.,
$\langle \mathbb{P}_{\mathrm{ext}}^{W}\rangle > \langle \mathbb{P}_{\mathrm{ext}}^{I}\rangle$ (Fig.~\ref{fig:heatmap_A_tau_all}d). This suggests that introductions through the environmental reservoir are less likely to establish than introductions through a single infected host. Furthermore, the magnitude of this difference depends on which transmission parameter varies seasonally. When $\gamma(t)$ or $c(t)$ is seasonal (Cases I and III), the difference between the two average extinction probabilities is relatively small and generally decreases as seasonal amplitude increases. In contrast, when $\kappa(t)$ varies seasonally (Case II), the difference is substantially more pronounced and generally increases with amplitude, indicating that seasonal variation in environment--host contact more strongly differentiates the invasion prospects of the two introduction pathways.

\subsection{Effect of parameter uncertainty on epidemiological outcomes} \label{sec:NR_GSA}

To determine which uncertain parameters most strongly influence outbreak behavior, we perform global sensitivity analysis (GSA) using the Saltelli sampling scheme and Jansen estimators introduced in Sec.~\ref{sec:GSA} and detailed in Appendix~\ref{appendix:GSA}. For the stochastic CTMC model, we estimate each quantity of interest (QoI) by averaging over repeated stochastic simulations at each sampled parameter set. The resulting sensitivity indices therefore quantify how uncertainty in model parameters affects expected outbreak outcomes.

\begin{figure}[H]
\centering
\includegraphics[width=0.7\textwidth]{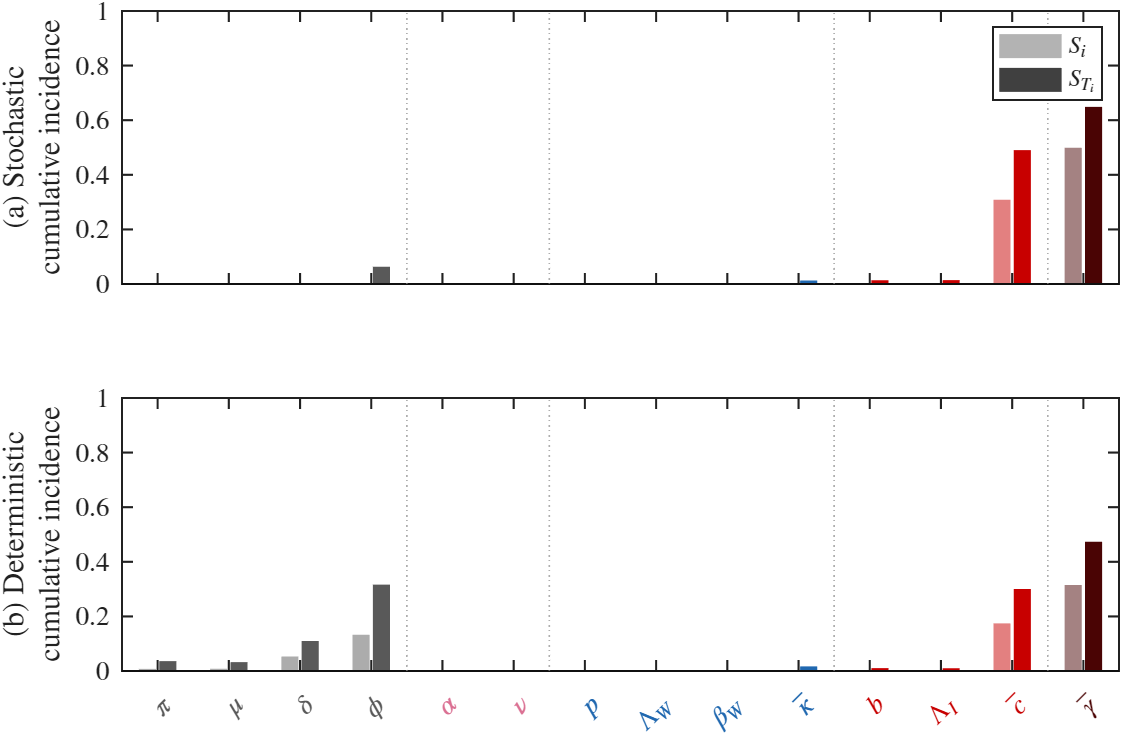} \caption{Global sensitivity analysis of cumulative incidence of infected hosts for the stochastic CTMC model and the deterministic model, with parameters sampled over the ranges listed in Table~\ref{SFDParamTable}. Bars show the first-order Sobol index $S_i$ and total-order Sobol index $S_{T_i}$ for each uncertain parameter. Parameter labels are colored by class: host demographic and disease progression parameters in gray, environmental transmission parameters in blue, and direct-transmission parameters in red. The stochastic analysis in (a) uses $N=500$ Saltelli samples and $K=100$ independent CTMC simulations per sampled parameter set, while the deterministic analysis in (b) uses $N=10,000$ Saltelli samples. All simulations use the baseline parameter values in
Table~\ref{SFDParamTable} with initial conditions
$S(0)=1000$, $I(0)=1$, and $W(0)=0$.}
\label{fig:GSA_stochastic}
\end{figure}

In both the stochastic and deterministic models, average host susceptibility $\bar{\gamma}$ and the average host--host contact rate $\bar{c}$ have the largest first-order and total-order effects, as determined by Sobol indicies (Fig. \ref{fig:GSA_stochastic}). This indicates that cumulative incidence is driven primarily by variation in host susceptibility and direct host--host transmission. The deterministic model also shows a noticeable contribution from parameters controlling infected host removal, particularly the recovery rate, $\phi$, and disease-induced mortality rate, $\delta$. As the deterministic model does not include stochastic fadeout, parameters controlling the duration of infection have a more direct effect on the cumulative outbreak trajectory. In the stochastic CTMC model, sensitivity is more dependent on the transmission parameters, reflecting the importance of early invasion and extinction when infected hosts and environmental pathogen loads are low.

The remaining parameters have negligible first-order and total-order effects for cumulative host incidence. This does not imply that these parameters are mechanistically unimportant. Instead, within the parameter ranges considered in our numerical simulations, variation in these parameters produces relatively small variation in cumulative host incidence compared with uncertainty in $\bar{\gamma}$, $\bar{c}$, and $\phi$. We observe broadly similar global sensitivity patterns for the other QoIs reported in Appendix~\ref{appendix:GSA_fig} (Figs. \ref{fig:GSA_stochastic_I_appen}, \ref{fig:GSA_stochastic_W_appen}, and \ref{fig:GSA_det_additional_det}). However, QoIs related to extinction probability show some differences, with parameters associated with environmental exposure becoming more influential for extinction probabilities following environmental introduction (Figs.~\ref{fig:GSA_stochastic_W_appen} and \ref{fig:GSA_det_additional_det}).

\section{Discussion} \label{sec:discussion}

Environmentally persistent pathogens can spread through both direct host--host transmission and indirect transmission from pathogen shed into the environment by infected hosts. This dual transmission structure creates multiple opportunities for stochastic fadeout, especially when introduction occurs during periods of low host susceptibility or weak contact with the relevant transmission pathway. In this study, we developed a stochastic seasonal SIW model to examine how dose--response, introduction route, and seasonal timing shape disease invasion and extinction dynamics. We used a multitype branching process approximation to estimate extinction probabilities following introduction by either a single infected host or a single unit of environmental contamination.

Using snake fungal disease as a motivating case study, we find that outbreak risk depends both on the magnitude of transmission parameters and on the timing of pathogen introduction relative to seasonal changes in host susceptibility, environmental exposure, and host--host contact. Environmental dose--response functions further shape outbreak timing and epidemic intensity when environmental pathogen loads are sufficiently high, with more rapid rising dose--response functions producing earlier epidemics and larger disease burdens. The route of introduction also affects extinction risk because pathogens introduced through environmental contamination may encounter different seasonal conditions from those introduced by an infected host. Seasonality therefore creates windows of high and low extinction probability, particularly when host susceptibility is not aligned with the active transmission pathway. Invasion and extinction dynamics are shaped most strongly by parameters controlling direct host transmission and infectious duration, including how readily susceptible hosts become infected, how often hosts contact one another, and how long infected hosts remain infectious.

These findings emphasize that the importance of  seasonality extends beyond the magnitude of seasonal variations or their annual averages. The relative timing of susceptibility, environmental exposure, and host--host contact also determines whether conditions favor pathogen establishment. Additionally, peak extinction probability often lags behind seasonal drivers (as seen in other systems \cite{hridoy2025investigating, carmona2020winter}) by an interval that changes with effective transmission intensity and the route of introduction, suggesting that interventions for environmentally persistent pathogens should begin before apparent high-risk periods, such as reducing environmental contamination ahead of seasonal increases in host activity.

This framework builds on stochastic model analyses by using a multitype branching process approximation to capture the dependence of extinction probability on introduction timing, introduction route, and seasonal synchrony. Branching process approximation is an efficient tool for identifying seasonal windows of invasion risk, particularly because CTMC simulations can be computationally expensive when extinction probabilities must be evaluated across many seasonal phases and parameter combinations. In addition, a useful feature of our global sensitivity analysis approach is that it separates uncertainty due to parameter variation from uncertainty due to stochastic simulation noise. 

Our work is not the first to address stochasticity in the context of environmental transmission. Notably, Lahodny et al. \cite{lahodny2015estimating} developed a stochastic CTMC formulation extending the deterministic framework of Bani-Yaghoub et al. \cite{bani2012effectiveness}. They approximated early invasion dynamics using a multitype branching process, allowing them to estimate extinction and major-outbreak probabilities. Using salmonellosis and cholera as case studies, they showed that shedding, environmental replication, indirect transmission, and decontamination strongly influence invasion risk, even when deterministic thresholds predict persistence. However, their parameters did not vary seasonally, and they assumed a fixed functional relationship between environmental pathogen load and infection. Both assumptions may substantially influence invasion and extinction risk in seasonal host--pathogen systems, as we have shown through the inclusion of seasonality in our stochastic seasonal framework.

In the current framework, seasonal behavior is incorporated through changes in susceptibility and contact rates, rather than through the underlying behavioral processes themselves. For example, changes in exposure duration, contact frequency, and contact duration are all folded into effective transmission terms. Future work could separate these components more explicitly and test how each one changes the timing and probability of disease establishment. Extending the model beyond a single host species would also make it possible to examine how differences in susceptibility, shedding, direct contact, or contact with contaminated environments among species affect stochastic extinction.

In the broader context of environmentally transmitted diseases, our results show that disease establishment depends on more than the average transmission intensity. The timing and route of introduction, the dose--response relationship, and the seasonal alignment between susceptibility and contact jointly shape whether an outbreak takes off or fades out. Accounting for these interactions makes stochastic extinction an integral part of the epidemiological story, especially for pathogens that can persist in the environment even when infected host numbers are low.

\newpage

\appendix
\setcounter{figure}{0}
\renewcommand{\thefigure}{S\arabic{figure}}  

\section{Seasonal parameters}\label{appendix:tau}

We incorporate seasonal forcing for susceptibility ($\gamma$) and contact rate ($\kappa$ and $c$) parameters. These take the form of positive, $\omega$-periodic functions of the form:
\begin{equation}\label{eq:generalseasonal}
\eta(t) \;=\; \bar{\eta}\Big(1 + A_{\eta}\,\sin\!\big(2\pi (t/\omega + \tau_{\eta})\big)\Big),
\quad 0 \le A_{\eta} \le 1,\;\; 0 \le \tau_{\eta} \le 1,
\end{equation}
where $\bar{\eta}$ denotes the mean value of the seasonal parameter $\eta(t)$, $A_{\eta}$ is the amplitude of periodic variation, and $\tau_{\eta}$ is the phase shift determining the timing of the seasonal peak within each period. 

Specifically, we model the environmental--host contact rate $\kappa(t)$ by
\begin{equation}\label{eq:kappa}
\kappa(t) \;=\; \bar{\kappa}\Big(1 + A_{\kappa}\,\sin\!\big(2\pi (t/\omega + \tau_{\kappa})\big)\Big),
\quad 0 \le A_{\kappa} \le 1,\;\; 0 \le \tau_{\kappa} \le 1,
\end{equation}

\noindent where $\bar{\kappa}$ is the mean environment--host  contact rate, $A_{\kappa}$ is the amplitude of periodic variation, and $\tau_{\kappa}$ determines the timing of peak exposure within each period. 

Similarly, we allow the contact rate $c(t)$ to vary periodically

\begin{equation}\label{eq:cij}
c(t)
\;=\;
\bar{c}
\Big(1 + A_{c}\,\sin\!\big(2\pi(t/\omega + \tau_{c})\big)\Big),
\quad
0 \le A_{c} \le 1,\;
0 \le \tau_{c} \le 1,
\end{equation}
where $\bar{c}$ is the mean contact level, $A_c$ is the amplitude of periodic variation, and $\tau_c$ specifies the timing of peak contact.

Motivated by empirical work demonstrating that the probability of successful infection
depends not only on pathogen dose but also on host susceptibility and temporal context
of exposure, we model susceptibility as a seasonal function, $\gamma(t)$, which
scales transmission for hosts

\begin{equation}\label{eq:gamma}
\gamma(t) \;=\; {\bar{\gamma}
\Big(} 1 + A_{\gamma} \,\sin\!\big(2\pi (t/\omega + \tau_{\gamma})\big)\Big),
\quad 0 \le A_{\gamma} < 1,\;\; 0 \le \tau_{\gamma} \le 1,
\end{equation}
where $\bar{\gamma}$ is the mean susceptibility, $A_{\gamma}$ is the amplitude of periodic variation, and $\tau_{\gamma}$ specifies the timing of peak susceptibility.

\begin{figure}[H]
\centering
\includegraphics[width=0.9\textwidth]{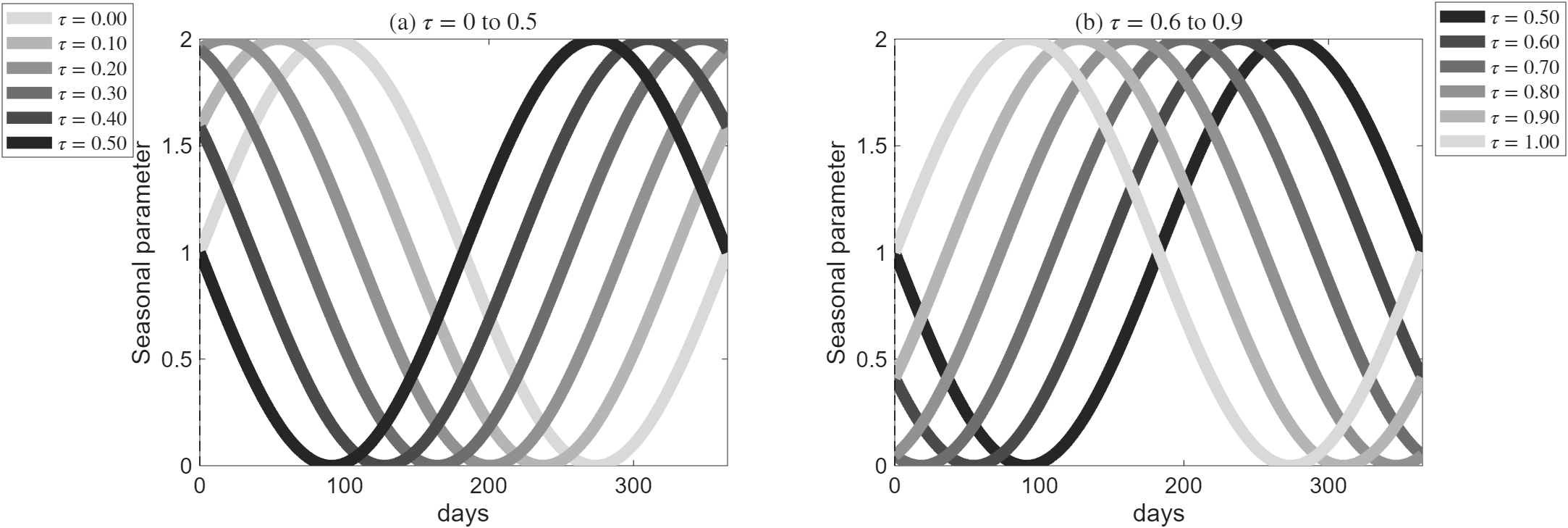} \caption{Seasonal parameter curves as the phase shift $\tau$ varies.}
\label{fig:seasonal_tau}
\end{figure}

To illustrate the role of the seasonal phase shift $\tau$, we consider a generic periodic parameter of the form
\[
\eta(t)=1+\sin\!\left(2\pi\left(\frac{t}{\omega}+\tau\right)\right),
\]
where $\omega$ denotes the period and $\tau$ is a phase shift controlling the timing of the seasonal peak. Varying $\tau$ shifts the curve horizontally without altering its amplitude or period. Figure~\ref{fig:seasonal_tau} illustrates this effect for several representative values of $\tau$.

\section{Derivation of basic reproduction numbers}\label{appendix:R0}

To compute $\mathcal{R}_0$, we linearize the SIW model around the DFE, $(S^*, I^*, W^*) = (\bar N, 0, 0).$
Near the DFE, the dose--response functions are approximated by their linearizations, $f_W(pW) \approx f_W'(0)\,pW, f_I(b I) \approx f_I'(0)\,bI,$ and the susceptible population is fixed at its disease-free value, $S(t)\approx S^*=\bar N$.
Using the next-generation matrix approach, the Jacobian of the linearized system (focusing just on infected states $I$ and $W$) is decomposed
into new infection terms $F(t)$ and transition terms $V$, so that
$J(t)=F(t)-V$ \cite{wang2008threshold, van2002reproduction}. The matrices $F(t)$ and $V$, evaluated at the DFE, are given by
\[
F(t)=
\begin{pmatrix}
\gamma(t)c(t) f_I'(0)b\,\bar N 
& \gamma(t)\kappa(t) f_W'(0)p\,\bar N \\[4pt]
\alpha & 0
\end{pmatrix},
\qquad
V=
\begin{pmatrix}
\mu+\delta+\phi & 0\\[2pt]
0 & \nu
\end{pmatrix}.
\]

\begin{enumerate}[(i)]
\item Constant environment:

In a constant environment, the periodic parameters are replaced by their average values,
i.e., $\kappa(t)\equiv\bar\kappa$, $c(t)\equiv\bar c$, and $\gamma(t)\equiv\bar\gamma$.
In this case, the matrices $F$ and $V$ are constant, and the basic reproduction number is given by
\[
\mathcal{R}_0^{\mathrm{constant}}=\rho(FV^{-1}).
\]
where \[
FV^{-1}=
\begin{pmatrix}
\dfrac{\bar\gamma \bar c f_I'(0)b\,\bar N}{\mu+\delta+\phi}
&
\dfrac{\bar\gamma \bar\kappa f_W'(0)p\,\bar N}{\nu}
\\[12pt]
\dfrac{\alpha}{\mu+\delta+\phi}
&
0
\end{pmatrix}.
\]

Equivalently, 
\begin{equation}\label{eq:R0_constant}
\mathcal R_0^{\mathrm{constant}}
=
\frac{1}{2}
\left[
\frac{\bar\gamma\bar c f_I'(0)b\bar N}{\mu+\delta+\phi}
+
\sqrt{
\left(
\frac{\bar\gamma\bar c f_I'(0)b\bar N}{\mu+\delta+\phi}
\right)^2
+
\frac{4\alpha\bar\gamma\bar\kappa f_W'(0)p\bar N}
{\nu(\mu+\delta+\phi)}
}
\right].
\end{equation}

The DFE $(\bar N,0,0)$ is locally asymptotically
stable if $\mathcal{R}_0^{\mathrm{constant}}<1$ and unstable if
$\mathcal{R}_0^{\mathrm{constant}}>1$ \cite{wang2008threshold, van2002reproduction}.

\item Seasonal environment:

In the seasonal environment, $F(t)$ is $\omega$--periodic due to seasonal variation in contact rates
and susceptibility, while $V$ remains constant. The basic reproduction number is therefore
computed using Floquet theory \cite{wang2008threshold, klausmeier2008floquet}.
Specifically, we solve the linear periodic system
\[
\dfrac{dX}{dt}=\big[F(t)-V\big]X,\quad X(0)=I_2,
\]
where $X= [I; W]$ is the system of infected states and $I_2$ is the $2\times 2$ identity matrix. The spectral radius of the monodromy matrix
$X(\omega)$ defines the basic reproduction number,
$\mathcal{R}_0^{\mathrm{per}}=\rho(X(\omega))$. An explicit closed form for $\mathcal{R}_0^{\mathrm{per}}$ is generally unavailable, so we compute
it numerically following the approach of Wang and Zhao \cite{wang2008threshold}. Introducing
a scalar parameter $\lambda>0$,
\begin{equation}\label{eq:floquet_SIW}
\dfrac{dX}{dt}=\left[\dfrac{F(t)}{\lambda}-V\right]X,\quad X(0)=I_2,
\end{equation}
we generate a sequence $\{\lambda_i\}$ until convergence to a value $\lambda_k$ satisfying
$|\rho(X(\omega,\lambda_k))-1|<10^{-6}$. The limiting value $\lambda_k$ provides a numerical
approximation of $\mathcal{R}_0^{\mathrm{per}}$. The DFE is stable if
$\mathcal{R}_0^{\mathrm{per}}<1$ and unstable if
$\mathcal{R}_0^{\mathrm{per}}>1$ \cite{wang2008threshold}.
\end{enumerate}

\section{Derivation of branching process approximation for the CTMC model}\label{appendix:BPA}

To derive the BPA, we linearize the system \eqref{eq:SIW} about the DFE, 
\[
(S^*, I^*, W^*) = (\bar N, 0, 0),
\]
where the susceptible class is fixed at its disease–free value $S(t)\approx \bar N$, where $\pi=\mu$. Under this assumption, only the infectious components $(I,W)$ evolve stochastically.

Near the DFE, the dose–response functions are linearized via first–order Taylor expansion,

\[
f_W(p W) \approx f_W'(0)\,pW,
\qquad
f_I(b I) \approx f_I'(0)\,b I.
\]

Under this linearization, only the events from Table~\ref{tab:SIWctmc} that change $(I,W)$ contribute to the branching approximation (events 3-10)
and define the transition probabilities of the SIW BPA.

Let 
\[
\mathbb{P}_{(i,w),(j,\ell)}(t_0,t)
=
P\!\Big(\big(I(t),W(t)\big)=(i,w)\,\big|\,\big(I(t_0),W(t_0)\big)=(j,\ell)\Big)
\]
denote the transition probability for the BPA.  
The initial states of the BPA correspond to: one infectious host, $(I=1,W=0)=\mathbf e_I$, or one unit of environmental pathogen, $(I=0,W=1)=\mathbf e_W.$
The absorbing state is disease extinction, $(I,W)=(0,0)$.

The time interval $[t_0,t]$ is partitioned into a short increment 
$[t_0,t_0+\Delta t_0]$ and the remaining interval $[t_0+\Delta t_0,t]$. 
During $[t_0,t_0+\Delta t_0]$ at most one event may occur, and all rates are evaluated at $t_0$.  
Denoting the current state by $(i,w)$, we obtain
\begin{align*}
\mathbb{P}_{(i,w),(j,\ell)}(t_0,t)
&= \underbrace{\gamma(t_0)\,\kappa(t_0) f_W'(0)\,p\,\bar N\, w\,\Delta t_0}_{i\to i+1}
   \mathbb{P}_{(i+1,w),(j,\ell)}(t_0+\Delta t_0,t) \\[3pt]
&\quad + \underbrace{\gamma(t_0)\,c(t_0) f_I'(0)\,b\,\bar N\, i\,\Delta t_0}_{i\to i+1}
   \mathbb{P}_{(i+1,w),(j,\ell)}(t_0+\Delta t_0,t) \\[3pt]
&\quad + \underbrace{(\mu+\delta+\phi)\,i\,\Delta t_0}_{i\to i-1}
   \mathbb{P}_{(i-1,w),(j,\ell)}(t_0+\Delta t_0,t) \\[3pt]
&\quad + \underbrace{\alpha i\,\Delta t_0}_{w\to w+1}
   \mathbb{P}_{(i,w+1),(j,\ell)}(t_0+\Delta t_0,t) \\[3pt]
&\quad + \underbrace{\nu w\,\Delta t_0}_{w\to w-1}
   \mathbb{P}_{(i,w-1),(j,\ell)}(t_0+\Delta t_0,t) \\[3pt]
&\quad + \Big[ \underbrace{1-\mathcal{N}(t_0)\Delta t_0}_{\text{no change}}\Big]
   \mathbb{P}_{(i,w),(j,\ell)}(t_0+\Delta t_0,t)
   + o(\Delta t_0),
\end{align*}
where the total event rate is
\[
\mathcal{N}(t_0)
= \gamma(t_0)\kappa(t_0)f_W'(0)\,p\,\bar N\, w
+ \gamma(t_0)c(t_0)f_I'(0)\,b\,\bar N\, i
+ (\mu+\delta+\phi)i
+ (\alpha i  + \nu w).
\]

Subtracting $\mathbb{P}_{(i,w),(j,\ell)}(t_0+\Delta t_0,t)$, dividing by $\Delta t_0$, and letting $\Delta t_0\to 0$ yields the backward Kolmogorov differential equation (BKDE):
\begin{equation}\label{eq:BKDE_SIW_full}
\begin{aligned}
-\,\frac{\partial \mathbb{P}_{(i,w),(j,\ell)}(t_0,t)}{\partial t_0}
&= \gamma(t_0)\kappa(t_0)f_W'(0)\,p\,\bar N\, w\,\mathbb{P}_{(i+1,w),(j,\ell)}(t_0,t) \\[2pt]
&\quad + \gamma(t_0)c(t_0)f_I'(0)\,b\,\bar N\, i\,\mathbb{P}_{(i+1,w),(j,\ell)}(t_0,t) \\[2pt]
&\quad + (\mu+\delta+\phi)i\,\mathbb{P}_{(i-1,w),(j,\ell)}(t_0,t) \\[2pt]
&\quad + \alpha i\,\mathbb{P}_{(i,w+1),(j,\ell)}(t_0,t) \\[2pt]
&\quad + \nu w\,\mathbb{P}_{(i,w-1),(j,\ell)}(t_0,t) \\[2pt]
&\quad - \mathcal{N}(t_0)\,\mathbb{P}_{(i,w),(j,\ell)}(t_0,t).
\end{aligned}
\end{equation}

Let
\[
\mathbb{P}_{\mathbf e_I}(t_0,t) := \mathbb{P}_{(1,0)}(t_0,t),
\qquad
\mathbb{P}_{\mathbf e_W}(t_0,t) := \mathbb{P}_{(0,1)}(t_0,t).
\]
To derive extinction probabilities, we use the branching assumptions:
\[
\mathbb{P}_{(2,0)} = (\mathbb{P}_{\mathbf e_I})^2,\qquad
\mathbb{P}_{(1,1)} = \mathbb{P}_{\mathbf e_I}\,\mathbb{P}_{\mathbf e_W},\qquad
\mathbb{P}_{(0,2)} = (\mathbb{P}_{\mathbf e_W})^2,\qquad
\mathbb{P}_{(0,0)} = 1.
\]

Applying these assumptions to \eqref{eq:BKDE_SIW_full} yields two coupled BKDEs:

\begin{align}\label{eq:BKDE_I_single}
-\,\frac{\partial \mathbb{P}_{\mathbf e_I}}{\partial t_0}
&=
A_I(t_0)\,\mathbb{P}_{\mathbf e_I}^{\,2}
\;+\;
\alpha\,\mathbb{P}_{\mathbf e_I}\mathbb{P}_{\mathbf e_W}
\;+\;
D_I
\;-\;
\big(A_I(t_0)+\alpha + D_I\big)\mathbb{P}_{\mathbf e_I}.
\end{align}

where

\[
A_I(t_0) = \gamma(t_0)c(t_0)f_I'(0)\,b\,\bar N,
\qquad
D_I = \mu+\delta+\phi.
\]

and, 

\begin{align}\label{eq:BKDE_W_single}
-\,\frac{\partial \mathbb{P}_{\mathbf e_W}}{\partial t_0}
&=
A_W(t_0)\,\mathbb{P}_{\mathbf e_I}\mathbb{P}_{\mathbf e_W}
\;+\;
\nu
\;-\;
\big(A_W(t_0)+\nu\big)\mathbb{P}_{\mathbf e_W}.
\end{align}

where,

\[
A_W(t_0) = \gamma(t_0)\kappa(t_0)f_W'(0)\,p\,\bar N.
\]

\section{Global Sensitivity Analysis Methodology}\label{appendix:GSA}

Let $x = (x_1,\dots,x_d)$ denote the uncertain model parameters, and let $Y^{(k)}(x)$ denote the model output from the $k^{\text{th}}$ simulation of the CTMC at a parameter set $x$. For each parameter set $x$, we perform $K$ independent simulations, producing outputs $Y^{(1)}(x), \dots, Y^{(K)}(x)$. For a given parameter set $x$, the QoI, denoted by $Q(x)$, is defined as
\begin{equation}
Q(x) = \mathbb E[Y(x)\mid x].
\end{equation} 

\noindent Since repeated simulations at the same parameter value may yield different outcomes, $Q(x)$ cannot be computed analytically. Therefore, we estimate it using the sample mean of $K$ independent CTMC simulations
\begin{equation}\label{eq:QoI}
\hat{Q}(x) = \frac{1}{K}\sum_{k=1}^K Y^{(k)}(x),
\end{equation}
which represents the average model behavior for a given parameter set and serves as a Monte Carlo estimator of  $Q(x)$. Throughout, approximated quantities are indicated with a hat  ($~ \hat{}~$) over the symbol.

\begin{enumerate}[(i)]
\item Sobol indices:

We aim to compute the first-order Sobol indices
\begin{equation}\label{eq:sobol_i}
S_i = \frac{\mathrm{Var}_{x_i}\!\left(\mathbb{E}[Q(x)\mid x_i]\right)}{\mathrm{Var}(Q(x))}
\end{equation}
and the total-order Sobol index
\begin{equation}\label{eq:sobol_tot}
S_{T_i} = 1 - \frac{\mathrm{Var}_{x_{\sim i}}\!\left(\mathbb{E}[Q(x)\mid x_{\sim i}]\right)}{\mathrm{Var}(Q(x))},
\end{equation}
where $x_{\sim i}$ denotes all parameters except $x_i$. These indices measure the direct and overall influence of each parameter, respectively. Larger values of $S_i$ and $S_{T_i}$ indicate greater influence of parameter $x_i$ on the QoI. However, we need to estimate these indices as described below.

\item Saltelli sampling scheme:

A direct evaluation of Sobol indices (Eqs.~\eqref{eq:sobol_i}--\eqref{eq:sobol_tot})  involves expectations nested within variances, which would require a prohibitively large number of model evaluations. To address this, we employ the Saltelli sampling scheme, which provides an efficient Monte Carlo approximation \cite{saltelli2008global}. 

Two independent sample matrices $A,B\in\mathbb{R}^{N\times d}$ are generated using a Sobol low-discrepancy sequence (implemented with {\tt sobolset} in Matlab), which produces quasi-random samples that more uniformly cover the unit hypercube $[0,1]^d$ compared to standard random sampling. Here, $N$ denotes the number of parameter samples used. These samples are then mapped to the parameter ranges using a linear transformation for each parameter. 

For each parameter $x_i$, a third matrix $C^{(i)}$ is constructed by replacing the $i$th column of $B$ with that of $A$, i.e.,
\begin{equation}
C^{(i)} = (B_1, \dots, B_{i-1}, A_i, B_{i+1}, \dots, B_d).
\end{equation}

For each row $j=1,\dots,N$, we evaluate the model at the parameter values specified by the $j^{\text{th}}$ rows of $A$, $B$, and $C^{(i)}$, obtaining $\hat{Q}_A^{(j)}$, $\hat{Q}_B^{(j)}$, and $\hat{Q}_{C^{(i)}}^{(j)}$. Each $\hat{Q}^{(j)}$ is computed as the average of $K$ independent stochastic simulations at the corresponding parameter set using Eq.~\eqref{eq:QoI}. Before accounting for repeated stochastic simulations, this reduces the computational cost from $\mathcal{O}(N^2)$ to $\mathcal{O}(N(d+2))$ \cite{saltelli2008global}.

\item Monte Carlo estimators:

The Sobol indices are approximated using Jansen's paired-difference estimators \cite{jansen1999analysis, puy2021irrigated}. These estimators are based on the same
Saltelli sampling design as the covariance-based estimators used in \cite{kornetzke2026quantifying}, but express the variance components in terms of paired squared differences. Although there exist various versions of the estimators, Jansen’s estimators are known to exhibit improved numerical stability and lower variance compared to alternative estimators, particularly for models with strong nonlinearity \cite{puy2022comprehensive}.


Let $\{ \hat{Q}_A^{(j)} \}_{j=1}^N$ denote the QoI estimates evaluated on matrix $A$,  where each $\hat{Q}_A^{(j)}$ is computed as in Eq.~\eqref{eq:QoI}. Here, $N$ denotes the number of parameter samples used. The total variance of the QoI is estimated as 

\begin{equation}\label{eq:Jansen}
\mathrm{Var}(\hat{Q}(x)) \approx \frac{1}{N-1}
\sum_{j=1}^N
\left(
\hat Q_A^{(j)}
-
\frac{1}{N}\sum_{\ell=1}^N \hat Q_A^{(\ell)}
\right)^2.
\end{equation}

Using Jansen's estimators, the Sobol indices are computed as
\begin{equation}
S_i \approx 
1 - \frac{1}{2N} \sum_{j=1}^N 
\frac{\left( \hat{Q}_A^{(j)} - \hat{Q}_{C^{(i)}}^{(j)} \right)^2}
{\mathrm{Var}(\hat{Q}(x))},
\end{equation}
\begin{equation}
S_{T_i} \approx 
\frac{1}{2N} \sum_{j=1}^N 
\frac{\left( \hat{Q}_B^{(j)} - \hat{Q}_{C^{(i)}}^{(j)} \right)^2}
{\mathrm{Var}(\hat{Q}(x))}.
\end{equation}

\item Variance decomposition:

Since $\hat{Q}(x)$ is a Monte Carlo estimator based on $K$ stochastic simulations, its variance contains contributions from both parameter uncertainty and stochastic simulation noise. Following \cite{kornetzke2026quantifying}, we decompose total variance of $\hat{Q}(x)$ as
\begin{equation}\label{eq:var_decom}
\mathrm{Var}(\hat{Q}(x)) 
= \underbrace{\sigma_x^2}_{\text{parameter uncertainty}} + \qquad \underbrace{\dfrac{1}{K}\mathbb{E}_x[\sigma_s^2(x)]}_{\text{stochastic simulation noise}}
\end{equation}
\noindent where $\sigma_x^2 = \mathrm{Var}(Q(x))$ denotes the variance of the QoI due to parameter uncertainty, and $\sigma_s^2(x)$ is the variance due to the stochastic simulation of the model at fixed $x$.  At each parameter $x$, the corresponding stochastic variance is estimated by \begin{equation}\label{eq:var_solver}
\hat{\sigma}_s^2(x) = \frac{1}{K-1}\sum_{k=1}^K \left(Y^{(k)}(x) - \hat{Q}(x)\right)^2
\end{equation}

\noindent which represents how much the individual simulations vary around their average, i.e., the intrinsic randomness of the stochastic model at fixed parameters. Using Eq.~\eqref{eq:var_decom}, variance due to parameter uncertainty is then estimated by
\begin{equation}\label{eq:correct_var}
\hat{\sigma}_x^2 \approx \mathrm{Var}(\hat{Q}(x)) - \dfrac{1}{K}\mathbb{E}_x[\hat{\sigma}_s^2(x)].
\end{equation}

\noindent \noindent This is an approximation since both $\mathrm{Var}(\hat{Q}(x))$ and $\mathbb{E}_x[\sigma_s^2(x)]$ are estimated using Monte Carlo sampling over a finite number of parameter samples $N$, and $\hat{\sigma}_s^2(x)$ itself is approximated using a finite number of stochastic realizations $K$. The corrected variance $\hat{\sigma}_x^2$ is then used in the Sobol estimators and ensures that the resulting sensitivity indices reflect variability driven by parameter uncertainty rather than stochastic simulation noise.

\item Corrected Sobol estimators:

Once the corrected variance due to parameter uncertainty $\hat{\sigma}_x^2$ has been estimated from Eq.~\eqref{eq:correct_var}, the Sobol indices are computed using the corrected Jansen estimators.

The first-order Sobol index is estimated as
\begin{equation}
S_i \approx
1 -
\dfrac{
\frac{1}{2N}\sum_{j=1}^N
\left[
\left(\hat{Q}_A^{(j)} - \hat{Q}_{C^{(i)}}^{(j)}\right)^2
-
\frac{\hat{\sigma}_{s,A}^{2,(j)} + \hat{\sigma}_{s,C^{(i)}}^{2,(j)}}{K}
\right]
}{
\hat{\sigma}_x^2
}.
\end{equation}

Similarly, the total-order Sobol index is estimated by
\begin{equation}
S_{T_i} \approx
\dfrac{
\frac{1}{2N}\sum_{j=1}^N
\left[
\left(\hat{Q}_B^{(j)} - \hat{Q}_{C^{(i)}}^{(j)}\right)^2
-
\frac{\hat{\sigma}_{s,B}^{2,(j)} + \hat{\sigma}_{s,C^{(i)}}^{2,(j)}}{K}
\right]
}{
\hat{\sigma}_x^2
}.
\end{equation}
\end{enumerate}

\section{Derivation of parameter values}\label{appendix:param}

Parameter values used in the numerical simulations are found in Table~\ref{SFDParamTable}. 
Here, we summarize how these values are derived. Parameter values labeled “Experiment” in Table \ref{SFDParamTable} are derived from unpublished experimental data. Ranges for these parameters are from the cited sources and encompass our derived parameter values.  

The host natural mortality rate $\mu$ is based on annual survival estimates for ratsnakes reported in \cite{weatherhead2012mortality}.
Mortality rates are obtained by converting survival estimates to daily hazards using $\mu = -\ln(S_r)/365$, where $S_r$ denotes annual survival.
Using the reported 95\% confidence interval for adult annual survival ($0.50$--$0.91$) yields $\mu \in (0.000258,0.001899)$ day$^{-1}$. 
The recruitment rate $\pi$ is chosen to balance natural mortality at the disease-free equilibrium, so that $\pi=\mu$ and the host population remains approximately constant in the absence of infection. 
The disease-induced mortality rate $\delta$ is set to $0$, consistent with field observations reporting no mortality
attributed to ophidiomycosis in some populations \citep{dillon2024effects}. 
To explore potential mortality scenarios we consider $\delta \in (0,0.0057)$ day$^{-1}$, derived from reports of approximately $40\%$ mortality over a $90$ day disease period \citep{woah_sfd}. 
Assuming a constant hazard, this corresponds to $\delta = -\ln(0.6)/90 \approx 0.0057$ day$^{-1}$. 
The recovery rate $\phi$ is based on experimental estimates of infection clearance. 
A mean clearance time of approximately $16.6$ days is reported which corresponds to $\phi = 1/16.6 \approx 0.0602$ day$^{-1}$.

The pathogen shedding rate $\alpha = 0.03$ day$^{-1}$ is chosen to be consistent with experimental observations of environmental contamination associated with infected snakes \citep{mckenzie2020ophidiomycosis}.
The environmental pathogen removal rate $\nu = 0.067$ day$^{-1}$ is chosen corresponding to an average environmental persistence time of approximately $15$ days \citep{campbell2021soil}. The parameters $p$ and $b$ scale environmental pathogen load and infected host prevalence into effective infection doses in the dose–response functions. Such parameters are commonly used in microbial dose–response modeling to relate pathogen exposure to infection probability \citep{haas2014quantitative}. We have chosen to have these be equal and set to one.

\section{Extinction analysis for environmental dose response formulations}\label{appendix:Extinction analysis}

We consider different dose response formulations for environmental transmission and report extinction probabilities estimated from 1000 CTMC sample paths and from the branching process approximation (BPA) together with $\mathcal{R}_0$ (Table~\ref{Tab:ExtinctionBW}) as initial environmental dose-response slope, $f_W'(0)=\Lambda_W$, increases. All other parameters are held constant at their baseline values given in Table~\ref{SFDParamTable}.

The effect of $\Lambda_W$ is weaker when the outbreak is initiated by a single infected host, with $\mathbb{P}_{\mathrm{ext}}^{I}$ changing only slightly as $\Lambda_W$ increases. In contrast, extinction probability following environmental introduction $\mathbb{P}_{\mathrm{ext}}^{W}$ decreases more strongly with $\Lambda_W$. At low environmental infectivity, extinction probabilities can be slightly higher for environmental introduction than for infected host introduction. However, as $\Lambda_W$ increases, $\mathbb{P}_{\mathrm{ext}}^{W}$ drops below $\mathbb{P}_{\mathrm{ext}}^{I}$. This suggests that environmental contamination can have higher invasion potential when small environmental doses are sufficiently infectious. Since the branching process approximation is based on the linearization around the disease-free equilibrium, its predictions depend only on $f_W'(0)=\Lambda_W$ and are therefore the same across dose-response forms. The close agreement between the CTMC estimates and BPA values suggests that the approximation captures the early stochastic invasion dynamics well.



\begin{table}[H]
\small
\caption{Comparison of extinction outcomes across environmental dose-response forms}
\label{Tab:ExtinctionBW}
\centering
\begin{tabular}{l c c c c c c}
\toprule
\textbf{Dose-response form} & \textbf{$\Lambda_W$} & \textbf{CTMC $\widehat{\mathbb{P}}^I_{\mathrm{ext}}$} & \textbf{CTMC $\widehat{\mathbb{P}}^W_{\mathrm{ext}}$} & \textbf{BPA $\mathbb{P}_{\mathrm{ext}}^{I}$} & \textbf{BPA $\mathbb{P}_{\mathrm{ext}}^{W}$} &$\mathcal{R}_0$ \\
\midrule
Linear & $1\;$ & $0.290\;$ & $0.307\;$ & $0.268\;$ & $0.314\;$ & $3.667\;$ \\
\quad & $2\;$ & $0.268\;$ & $0.184\;$ & $0.262\;$ & $0.185\;$ & $4.000\;$ \\
\quad & $3\;$ & $0.271\;$ & $0.122\;$ & $0.260\;$ & $0.131\;$ & $4.291\;$ \\
\quad & $4\;$ & $0.255\;$ & $0.097\;$ & $0.259\;$ & $0.102\;$ & $4.553\;$ \\
\quad & $5\;$ & $0.266\;$ & $0.066\;$ & $0.258\;$ & $0.083\;$ & $4.794\;$ \\
\midrule
Exponential
 & $1\;$ & $0.271\;$ & $0.326\;$ & \multicolumn{3}{c}{\multirow{4}{*}{Same as linearization}} \\
\quad & $2\;$ & $0.262\;$ & $0.182\;$ & \multicolumn{3}{c}{} \\
\quad & $3\;$ & $0.244\;$ & $0.131\;$ & \multicolumn{3}{c}{}  \\
\quad & $4\;$ & $0.241\;$ & $0.094\;$ & \multicolumn{3}{c}{}  \\
\quad & $5\;$ & $0.246\;$ & $0.098\;$ & \multicolumn{3}{c}{}  \\
\midrule
Beta-Poisson (approx.) 
  & $1\;$ & $0.256\;$ & $0.313\;$ & \multicolumn{3}{c}{\multirow{4}{*}{Same as linearization}} \\
\quad & $2\;$ & $0.247\;$ & $0.186\;$  & \multicolumn{3}{c}{} \\
\quad & $3\;$ & $0.241\;$ & $0.128\;$ & \multicolumn{3}{c}{} \\
\quad & $4\;$ & $0.257\;$ & $0.100\;$ & \multicolumn{3}{c}{} \\
\quad & $5\;$ & $0.237\;$ & $0.070\;$ & \multicolumn{3}{c}{} \\
\bottomrule
\end{tabular}
\end{table}


\section{Extinction analysis for variation in phase shift and introduction time}\label{appendix:intro_shift}

\begin{figure}[H]
\centering
\includegraphics[width=0.8\textwidth]{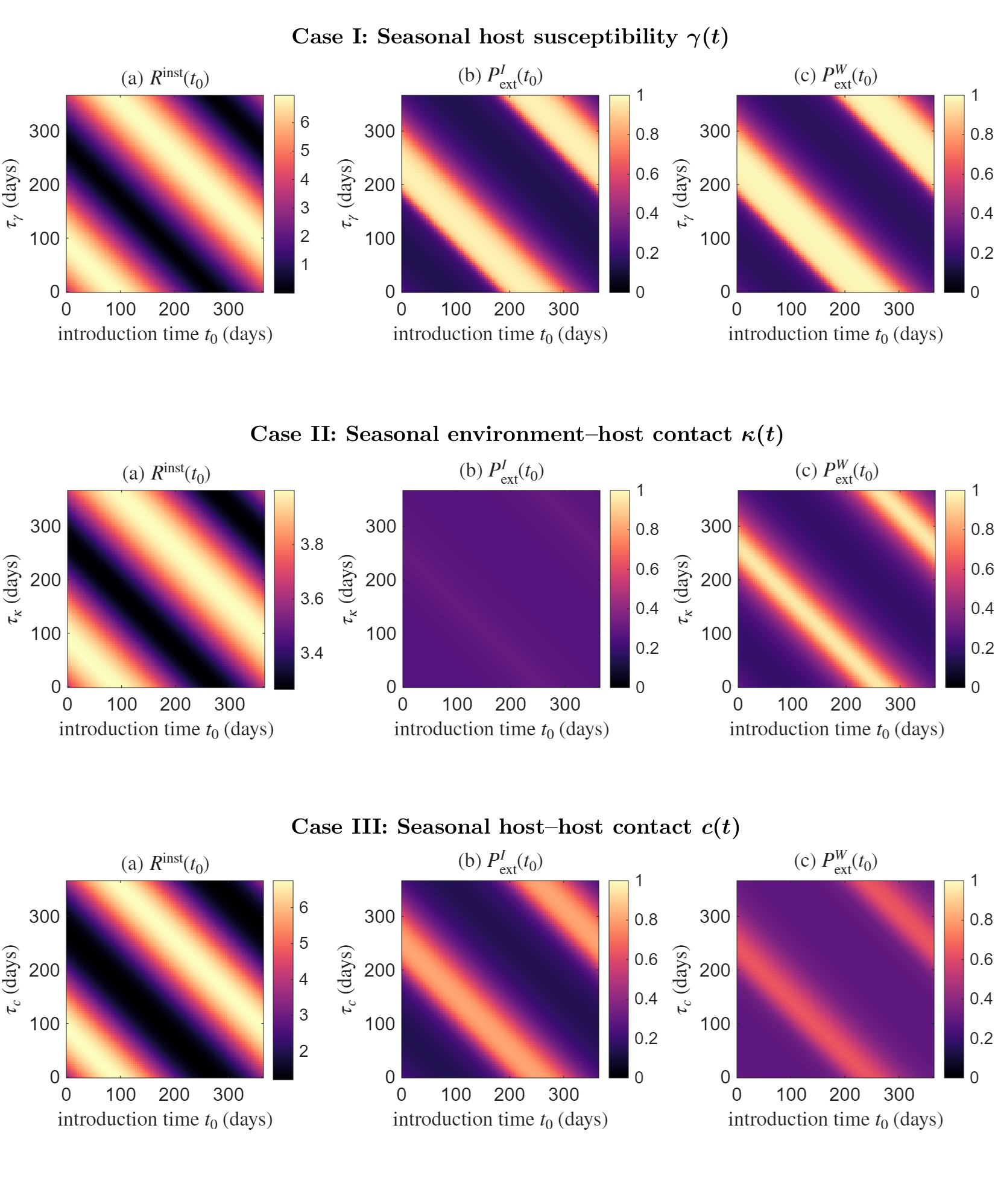} \caption{Columns show (a) the instantaneous reproduction number $\mathcal{R}^{\mathrm{inst}}(t_0)$, (b) $\mathbb{P}_{\mathrm{ext}}^{I}(t_0)$ for introductions initiated by a single infected host, and 
(c) $\mathbb{P}_{\mathrm{ext}}^{W}(t_0)$ for introductions initiated by a single environmental pathogen from the branching process approximation. Case I: seasonal variation in host susceptibility $\gamma(t)$, 
Case II: seasonal variation in host susceptibility $\gamma(t)$ environmental contact $\kappa(t)$, and 
Case III: seasonal variation in host--host contact $c(t)$.}
\label{fig:heatmap_tau_intro_all}
\end{figure}

Figure~\ref{fig:heatmap_tau_intro_all} shows the instantaneous reproduction number $\mathcal{R}^{\mathrm{inst}}(t_0)$ together with the extinction probabilities $\mathbb{P}_{\mathrm{ext}}^{I}(t_0)$ and $\mathbb{P}_{\mathrm{ext}}^{W}(t_0)$ obtained from the branching process approximation, while varying the phase shift $\tau_i$ ($i \in \{\gamma,\kappa,c\}$) and the introduction time $t_0$, corresponding to the introduction of either a single infectious host or a single environmental pathogen.The diagonal bands reflect that both $\tau_i$ and the introduction time $t_0$ jointly determine the point in the seasonal cycle at which pathogen introduction occurs. Increasing $\tau_i$ shifts the seasonal pattern of the parameters in time, so similar invasion outcomes occur along diagonals where the effective point in the seasonal curve at introduction is unchanged. The pattern of $\mathcal{R}^{\mathrm{inst}}(t_0)$ varies almost identically across all three cases, with differences arising in magnitude rather than overall structure (Fig.~\ref{fig:heatmap_tau_intro_all}a).

The extinction probabilities differ across the three seasonal drivers. When susceptibility $\gamma(t)$ varies seasonally (case I), the patterns in $\mathcal{R}^{\mathrm{inst}}(t_0)$, $\mathbb{P}_{\mathrm{ext}}^{I}(t_0)$, and $\mathbb{P}_{\mathrm{ext}}^{W}(t_0)$ are broadly similar (but inverted between $\mathcal{R}^{\mathrm{inst}}(t_0)$ and the probabilities of extinction), consistent with host susceptibility modulating both direct and environmental transmission pathways simultaneously. When environment--host contact rate $\kappa(t)$ varies seasonally (case II), the effect on $\mathbb{P}_{\mathrm{ext}}^{I}(t_0)$ is comparatively weak, whereas $\mathbb{P}_{\mathrm{ext}}^{W}(t_0)$ exhibits pronounced seasonal bands, since environment-to-host contact directly affects environmental introductions. In contrast, when $c(t)$ varies seasonally (case III), the stronger variation occurs in $\mathbb{P}_{\mathrm{ext}}^{I}(t_0)$, while changes in $\mathbb{P}_{\mathrm{ext}}^{W}(t_0)$ are more moderate because the effect of host-to-host contact on environmental introductions is indirect and mediated through subsequent host transmission.

\section{Global sensitivity analysis for additional quantities of interest} \label{appendix:GSA_fig}

\begin{figure}[H]
\centering
\includegraphics[width=0.8\textwidth]{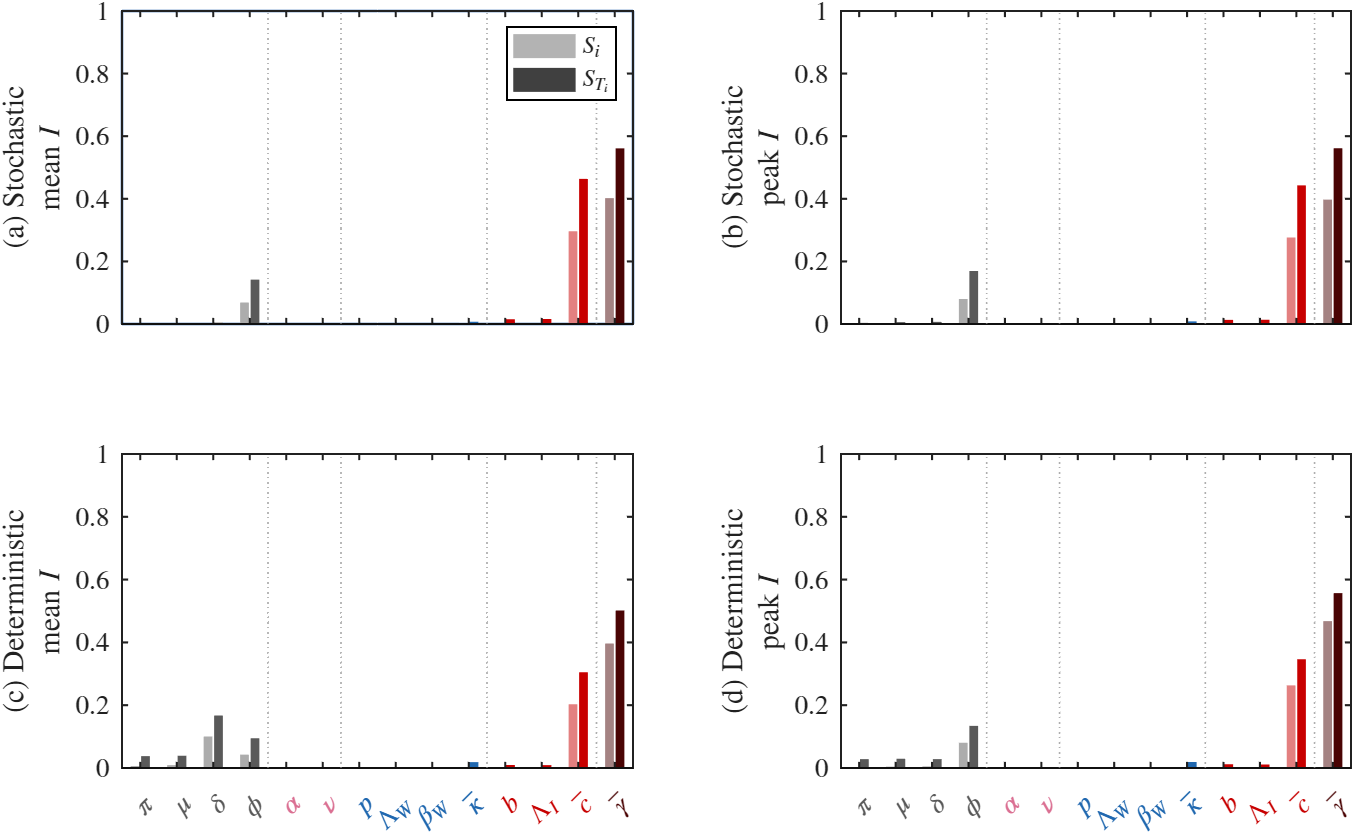} \caption{Global sensitivity analysis of mean and peak infected-host abundance for the stochastic CTMC model and the deterministic model. Bars show the first-order Sobol index $S_i$ and total-order Sobol index $S_{T_i}$ for each uncertain parameter, with parameters sampled over the ranges listed in Table~\ref{SFDParamTable}. The stochastic panels show sensitivities for (a) mean $I$ and (b) peak $I$, computed using $N=500$ Saltelli samples and $K=100$ independent CTMC simulations per sampled parameter set. The deterministic panels show sensitivities for (c) mean $I$ and (d) peak $I$, computed using $N=10,000$ Saltelli samples. Parameter labels are colored by parameter class. All simulations use the baseline parameter values in Table~\ref{SFDParamTable} with initial conditions $S(0)=1000$, $I(0)=1$, and $W(0)=0$.}
\label{fig:GSA_stochastic_I_appen}
\end{figure}

Figure~\ref{fig:GSA_stochastic_I_appen} shows that the sensitivity patterns for mean and peak infected host abundance are broadly consistent with those observed for cumulative incidence (Fig.~\ref{fig:GSA_stochastic}), with average host susceptibility $\bar{\gamma}$, average host--host contact rate $\bar{c}$, and host recovery rate $\phi$ having the largest influence in both the stochastic and deterministic models.

The sensitivity patterns for mean and peak environmental pathogen abundance are also broadly similar with the infected host QoIs. The average host susceptibility $\bar{\gamma}$ and average host--host contact rate $\bar{c}$ remain dominant, reflecting the role of infected hosts in seeding the environmental reservoir. In addition, parameters directly controlling environmental pathogen accumulation and clearance, particularly the shedding rate $\alpha$ and pathogen removal rate $\nu$, show noticeable sensitivity for environmental pathogen abundance.

\begin{figure}[H]
\centering
\includegraphics[width=0.85\textwidth]{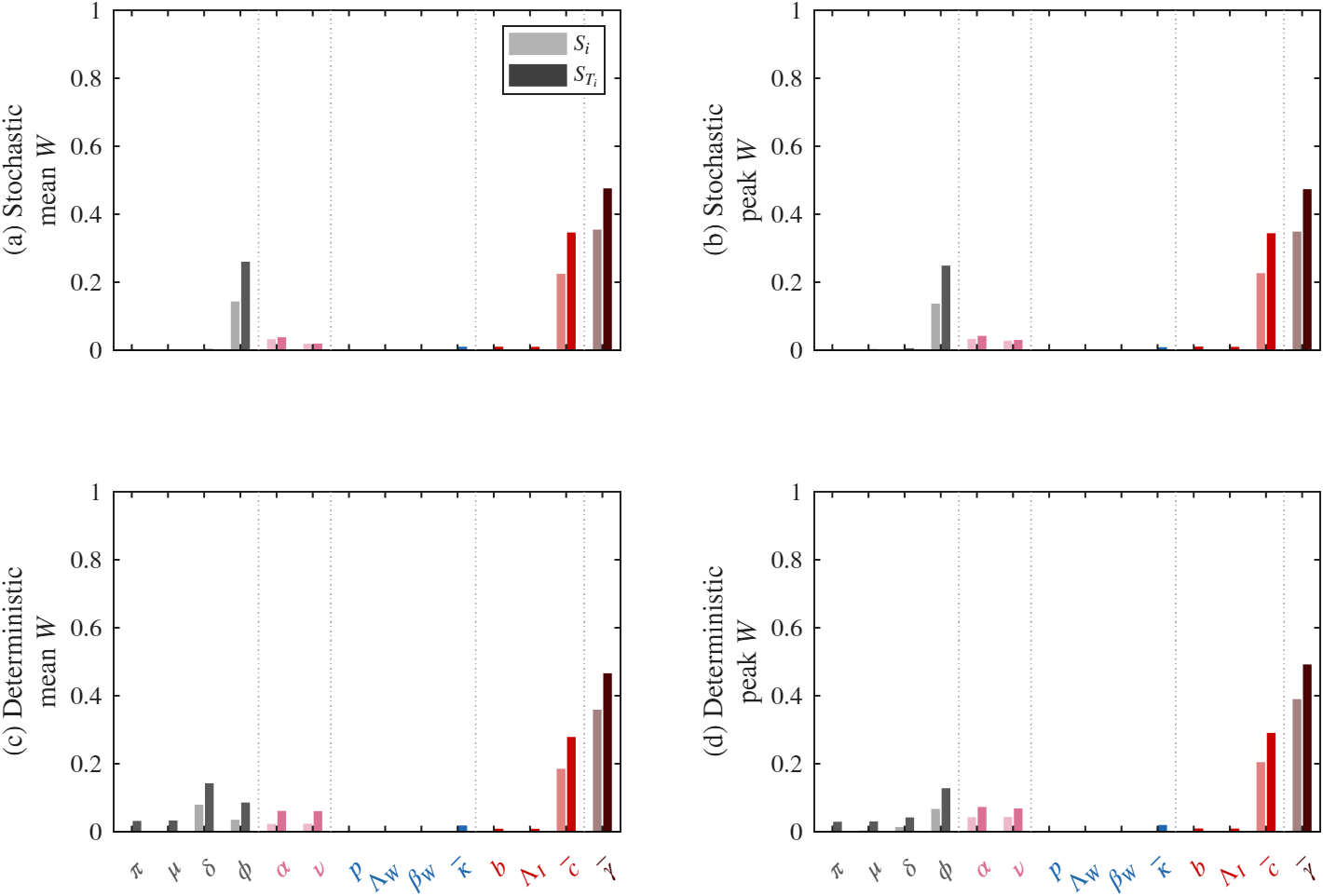} \caption{Global sensitivity analysis of mean and peak environmental pathogen abundance for the stochastic CTMC model and the deterministic model. Bars show the first-order Sobol index $S_i$ and total-order Sobol index $S_{T_i}$ for each uncertain parameter, with parameters sampled over the ranges listed in Table~\ref{SFDParamTable}. The stochastic panels show sensitivities for (a) mean $W$ and (b) peak $W$, computed using $N=500$ Saltelli samples and $K=100$ independent CTMC simulations per sampled parameter set. The deterministic panels show sensitivities for (c) mean $W$ and (d) peak $W$, computed using $N=10,000$ Saltelli samples. Parameter labels are colored by parameter class. All simulations use the baseline parameter values in Table~\ref{SFDParamTable} with initial conditions $S(0)=1000$, $I(0)=1$, and $W(0)=0$.}
\label{fig:GSA_stochastic_W_appen}
\end{figure}

\begin{figure}[H]
\centering
\includegraphics[width=0.9\textwidth]{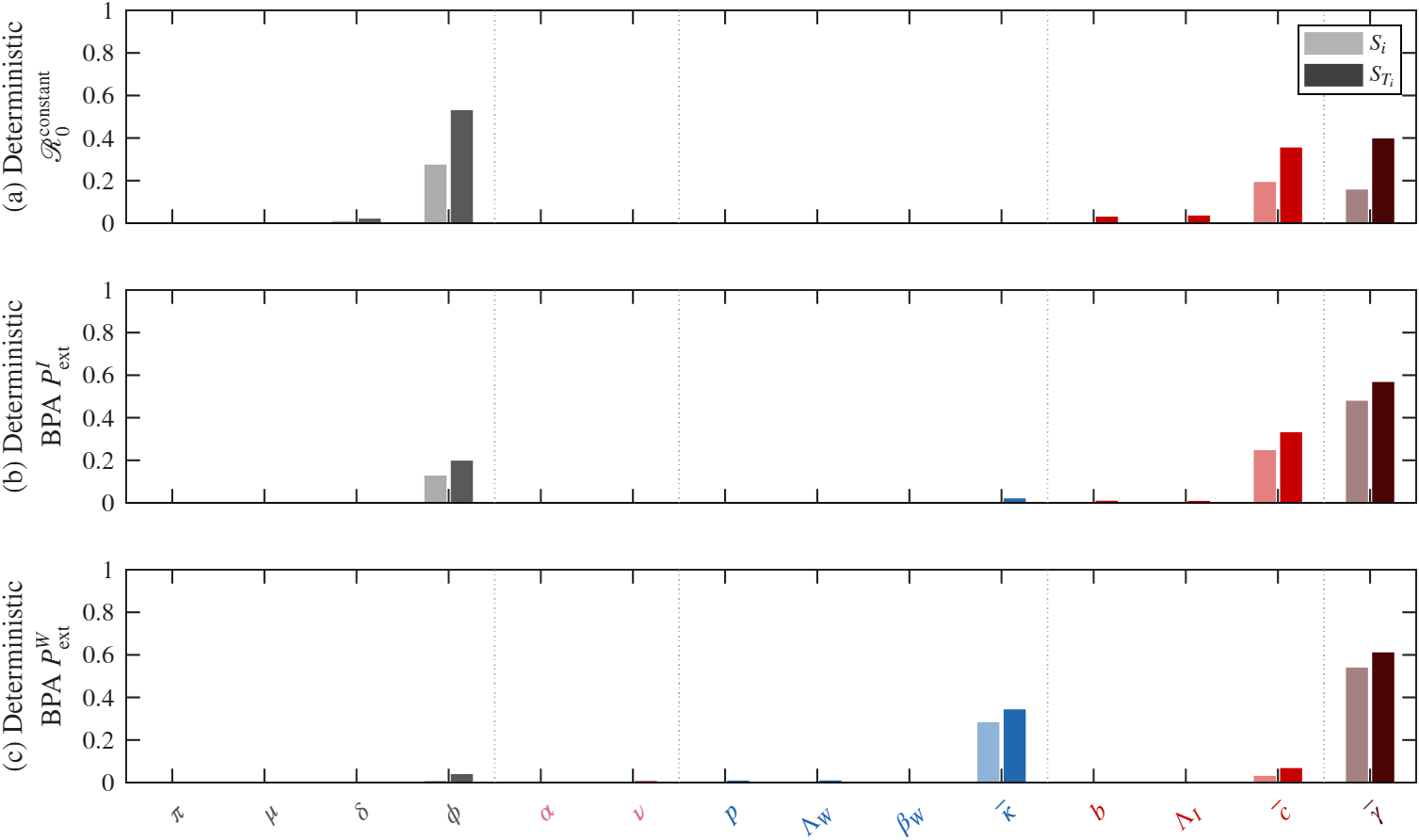} \caption{Global sensitivity analysis for deterministic quantities of interest. Panels show first-order Sobol indices $S_i$ and total-order Sobol indices $S_{T_i}$ for (a) basic reproduction number $\mathcal{R}_0^{\mathrm{constant}}$, (b) BPA extinction probability following introduction by a single infected host $\mathbb{P}_{\mathrm{ext}}^{I}$, and (c) BPA extinction probability following introduction by a single environmental pathogen unit $\mathbb{P}_{\mathrm{ext}}^{W}$. Parameters were sampled over the ranges listed in Table~\ref{SFDParamTable}. The analysis used $N=10,000$ Saltelli samples. Parameter labels are colored by parameter class.}
\label{fig:GSA_det_additional_det}
\end{figure}

As a complementary analysis, Figure~\ref{fig:GSA_det_additional_det} shows GSA results for deterministic QoIs: $\mathcal{R}_0^{\mathrm{constant}}$, $\mathbb{P}_{\mathrm{ext}}^{I}$, and $\mathbb{P}_{\mathrm{ext}}^{W}$. Across these deterministic QoIs, uncertainty in model outcomes is largely driven by average host susceptibility $\bar{\gamma}$, the average host--host contact rate $\bar{c}$, and the recovery rate $\phi$, consistent with the dominant patterns observed for cumulative incidence. However, the relative importance of these parameters depends on the QoI. The basic reproduction number $\mathcal{R}_0^{\mathrm{constant}}$ is most strongly influenced by the recovery rate $\phi$, followed by $\bar{\gamma}$ and $\bar{c}$. The extinction probability following introduction by a single infected host, $\mathbb{P}_{\mathrm{ext}}^{I}$, is mainly influenced by $\bar{\gamma}$ and $\bar{c}$, with a smaller contribution from $\phi$. In contrast, the extinction probability following introduction by a single environmental pathogen unit, $\mathbb{P}_{\mathrm{ext}}^{W}$, is most strongly influenced by $\bar{\gamma}$ and the mean environment--host contact rate $\bar{\kappa}$, reflecting the importance of environmental transmission when infection is introduced through the environmental reservoir.

\section*{Acknowledgments}

This work was supported by the National Institutes of Health–National Science Foundation (NIH-NSF) Ecology and Evolution of Infectious Disease award 1R01GM152978.

{\small
\setlength{\bibsep}{0pt}
\bibliographystyle{unsrt}
\bibliography{bibfile}
}

\end{document}